\documentclass[usenatbib,usegraphicx,oneside]{mn2e}
\usepackage{aas_macros}
\usepackage{times} 
\newcommand\arcs{\mbox{$^{\prime\prime}$}}%
\newcommand\arcm{\mbox{$^\prime$}}%
\def\lsim{~\raise0.3ex\hbox{$<$}\kern-0.75em{\lower0.65ex\hbox{$\sim$}}~}
\def\gsim{~\raise0.3ex\hbox{$>$}\kern-0.75em{\lower0.65ex\hbox{$\sim$}}~}

\newcommand{\chandra}{\textsl{Chandra}}
\newcommand{\hst}{\textsl{HST}}
\newcommand{\spitzer}{\textsl{Spitzer}}

\begin{document}
\title[The HDF-N SCUBA Super-map III]
{The HDF-North SCUBA Super-map III: Optical and near-infrared properties of submillimetre galaxies}

\author[Pope et. al.]{
\parbox[t]{\textwidth}{
\vspace{-1.0cm}
Alexandra Pope$^{1}$,
Colin Borys$^{2}$,
Douglas Scott$^{1}$,
Christopher Conselice$^{2}$,
Mark Dickinson$^{3,4}$,
Bahram Mobasher$^{4}$
}
\vspace*{6pt}\\
$^{1}$ Department of Physics \& Astronomy, University of British Columbia,
       Vancouver, BC, V6T 1Z1, Canada \\
$^{2}$ California Institute of Technology, Pasadena, CA 91125, USA\\
$^{3}$ National Optical Astronomy Observatory, Tucson, AZ 85719, USA\\
$^{4}$ Space Telescope Science Institute, Baltimore, MD, 21218, USA\\
\vspace*{-0.5cm}}

\date{Accepted for publication in MNRAS 2005}

\maketitle

\begin{abstract}
We present a new sub-mm `Super-map' in the HDF-North region (GOODS-North field), containing 40 statistically robust sources at 850$\,\mu$m. This map contains additional data, and several new sources, including one of the brightest blank-sky extragalactic sub-mm sources ever detected. We have used the ACS \emph{HST} images and ground-based near-IR observations from GOODS, along with deep radio observations, to develop a systematic approach for counterpart identification. With the depth achieved by this survey, optical counterparts have been found for all the radio-detected sub-mm sources. We have used the colours, morphologies and photometric redshifts of these secure identifications to help identify counterparts to the radio-undetected sources, finding that certain combinations of optical properties can be used to successfully identify the counterpart to a sub-mm source. 72 per cent of our sources with optical coverage have a unique optical counterpart using our new techniques for counterpart identification, and an additional 18 per cent have more than one possibility that meet our criteria in the ACS images. Thus only $\sim10$ per cent of our sources lack a plausible optical/near-IR counterpart, meaning that we have the first sample of SCUBA sources which is nearly completely identified in the optical. We have found a much higher ERO rate than other sub-mm surveys, due to the increased depth in the optical images. The median photometric redshift (and quartile range), from optical and near-infrared data, is 1.7 (1.3--2.5) for the radio-detected sub-mm sources, and rises to 2.3 (1.3--2.7) for the radio-undetected sub-sample. We find interesting correlations between the 850$\,\mu$m flux and both the $i_{\rm{775}}$ magnitude and the photometric redshift, from which there appears to be an absence of high redshift faint counterparts to the lower flux density SCUBA sources. While the quantitative morphologies span a range of values, in general the sub-mm galaxies show larger sizes and a higher degree of asymmetry than other galaxy populations at the same redshifts. In the appendix, we discuss several improvements in our data analysis procedure, including methods of testing for source reliability.
\end{abstract}


\section{Introduction}
Extragalactic submillimetre (sub-mm) surveys have revealed a population of high redshift galaxies that appear similar to nearby ultra-luminous infrared galaxies~\citep[ULIRG, e.g.][and references therein]{Blain02}. However, at high redshift we observe these galaxies at a time when they are thousands of times more numerous than such galaxies today~\citep[e.g.][]{Sanders99}. Therefore they play a significant role in galaxy formation and evolution and are thought to be the progenitors of present-day massive elliptical galaxies~\citep{Lilly99,8mJy_paperI}, although there is only very indirect evidence for this hypothesis~\citep[e.g.][]{Frayer99,Blain04}. The Sub-millimetre Common User Bolometer Array (SCUBA, Holland et al.~1999) on the James Clerk Maxwell Telescope (JCMT) has been used to find approximately 400 such objects since 1997, but the exact properties of these galaxies are still poorly understood. Progress in understanding this population is made by studying the characteristics of individual sources at other wavelengths. However, this is challenging, due to the JCMT beam size which creates a large search radius when looking for counterparts to the sub-mm emission at other wavelengths. Currently, coincidence with a $1.4\,$GHz radio source is the most successful way to identify the counterpart and refine the position of a SCUBA source (Ivison et al.~2000; Smail et al.~2000; Barger, Cowie, \& Richards 2000; Ivison et al.~2002; Smail et al.~2002; Borys et al.~2004; Clements et al.~2004). Studies that exploit this technique have found a mean spectroscopic redshift for the radio-detected sub-sample of SCUBA sources of 2.4~\citep{Chapman_nature}. However, SCUBA sources detected at $1.4\,$GHz represent only about half of the total number of sources found in all extragalactic SCUBA surveys (perhaps rising to $2/3$ for the brightest ones). Therefore there is a substantial fraction of SCUBA sources about which we currently know very little. Because the presence of a radio counterpart is sufficiently rare, it can be partially used to assess the reliability of sources in sub-mm surveys, and as a result doubt has been cast in the reality of some of the radio-undetected sub-mm sources~\citep{Ivison02,Greve04}. However, as we will demonstrate in this paper, in the region of our survey (where there is a wealth of both sub-mm and radio data), there are certainly a number of robust SCUBA sources with no radio counterpart, and we are thus able to compare the properties of the radio-detected and radio-undetected sub-samples.

In the absence of a radio-counterpart, we are forced to rely on shorter wavelength data, such as optical and infrared images. At optical wavelengths, the light from the energetic source in SCUBA galaxies will be heavily obscured by the dust that is producing the mid/far-IR emission. Combining this with the large beam size in typical sub-mm observations, it becomes clear that finding the correct counterpart is a challenge. Previous sub-mm surveys have found secure optical counterparts to less than half of their SCUBA sources; these tend to be faint and have red colours (Ivison et al.~2002; Webb et al.~2003; Clements et al.~2004), although these properties alone are not sufficient to pick out the correct optical counterpart~\citep{PaperII, Webb_ERO}. Such counterparts are often near the detection limit and so the deepest data at red and near-IR wavelengths are required to maximize the likelihood of detections.

The Great Observatories Origins Deep Survey~\citep[GOODS,][]{Giavalisco04} is a huge multi-wavelength campaign to study galaxy evolution in the early Universe. As part of the project, the Advanced Camera for Surveys (ACS) on the \textsl{Hubble Space Telescope} (\hst) has observed two large fields to produce deep high-resolution optical images and has detected galaxies down to $i_{\rm{AB}}\simeq28$ (Giavalisco et al.~2004). Deep \chandra\ data already exist for these fields, while observations with the \spitzer\ Space Telescope are also part of GOODS and were completed in November 2004. GOODS is therefore an ideal data--set for studying sub-mm sources.

In this paper we study the optical and near-IR properties of a large sample of SCUBA galaxies in the GOODS-North region, centred on the Hubble Deep Field North (HDF-N), in order to understand both the radio-detected and radio-undetected sub-samples. We use the new \hst\ observations, as well as ground-based near-IR imaging. A study of the \spitzer\ observations of our sub-mm sample will be part of a future paper. We include additional sub-mm observations of the region to update the `super-map' presented in Borys et al. (2003, hereafter Paper I) and include a revised source list. We discuss how the newer ACS data compare with the source identifications discussed in our earlier multi-wavelength study~\citep[hereafter Paper II]{PaperII}.

The format of this paper is as follows. Sections 2 and 3 describe the sub-mm, optical and near-IR data used in this paper, with the bulk of the sub-mm analysis presented in Appendix A. Section 4 explains how we calculated the parameters used to compare the optical sources. Our new technique for identifying counterparts is discussed in Section 5. Section 6 discusses the optical properties found for our sub-mm sample and Section 7 gives some conclusions. We have also included Appendix B, which describes the statistical methods we used to evaluate the robustness of our sub-mm sources. 

All magnitudes in this paper use the AB system unless otherwise noted. We assume a standard cosmology with $H_{0}=72\,\rm{km}\,\rm{s}^{-1}\,\rm{Mpc}^{-1}$, $\Omega_{\rm{M}}=0.3$ and $\Omega_{\Lambda}=0.7$.

\section{Submillimetre sample}
Currently, the largest amount of blank-field SCUBA data in a single field is found in the HDF-N region (Paper I). This unbiased survey, referred to as the `super-map', combines all SCUBA observations of the field taken by a number of different groups~\citep{Hughes98,Barger00,Borys02,Serjeant_hdf}. It is important to remember that SCUBA data always involve chopping and that the GOODS-North SCUBA data-set is composed of observations taken in all 3 SCUBA modes (photometry, jiggle-mapping and scan-mapping) with many different chop patterns, mostly using in-field chopping (i.e. with a chop throw that is less than the array size). For these reasons, when we construct the combined signal-to-noise SCUBA map, we refer to it as the `super-map', to emphasize that it is really telling us the best estimate for the signal-to-noise ratio for a point source centred on each pixel. This map and source list are described in Paper I, with an update and possible multi-wavelength counterparts discussed in Paper II. We have collected more SCUBA data in this region in an attempt to cover the entire GOODS-North field. New submillimetre observations and an improved source list are described in Appendix A, along with some changes to the data reduction over what is described in Paper I.  

The new 850$\,\mu$m `super-map' reveals a sample of 22 objects at $>4\sigma$ with an additional 18 at  $3.5$--$4.0\sigma$ (see Table~\ref{tab:cat850}), all but one of which overlap with the deep $HST$ observations from GOODS. For simplicity within this paper, we have assigned each of the 40 sources with a GOODS-North (GN) identification number. However, the full sub-mm source names (of the form SMMJ123...+62....) are listed in Table~\ref{tab:cat850} for future reference to these sources. One of our new sources, GN20, is extremely bright at 850$\,\mu$m and undetected in the radio. This source, along with the other new sources, are discussed in Section A4. 

In response to the recent claims that radio-undetected SCUBA sources may be spurious~\citep{Greve04}, we have performed several $\chi^{2}$ tests on the raw sub-mm data to search for some indication of this. As mentioned in Greve et al.~(2004), the lack of radio flux could be because these sources are simply at higher redshift, have cooler temperatures, or have more unusual SEDs than expected. It could also be due to different radio depths or u-v coverage; the VLA data in the HDF region reach a $1\sigma$ sensitivity of $7.5\mu$Jy (Richards 2000), which makes it one of the deepest images available. The results of both temporal and spatial $\chi^{2}$ tests reveal that there is no reason to distrust {\it any} of our sub-mm sources, and, in particular, the statistics on the radio-detected sources are no different from those of the radio-undetected sources. 

A recent paper by~\cite{Wang04} present an alternate list of sources in the HDF-N. While the two source lists have no major discrepancies (see appendix B of Paper II), our data analysis approach allows us to include more of the available data, and we also believe that our careful source extraction techniques provide better estimates for the sources in this region. Counterpart identification of sub-mm sources is also difficult and we have tried in this paper to rigorously follow a well-defined set of identification criteria, which leads to several differences compared with identifications previously suggested in~\cite{Wang04} and elsewhere.

\section{Multi-wavelength observations}
The HDF-N is one of the most extensively studied regions of the sky, with deep data existing across all wavebands.  Radio observations have been taken with the VLA over the whole super-map area at $1.4\,$GHz, and over a smaller region at $8.5\,$GHz~\citep{Richards00, Richards98}. $1.4\,$GHz observations have also been made with the WSRT over the whole field~\citep{Garrett00}, and with MERLIN over a smaller region~\citep{Muxlow99}. The \chandra\ 2 Msec image~\citep{Alexander03} provides the deepest X-ray survey of any part of the sky. Multi-wavelength properties of the Paper I source list, using primarily the radio and X-ray data, are presented in Paper II. Optical and near-IR data in Paper II were from the~\cite{Capak04} ground-based survey. However, we now have the ACS data, which provide deeper images and better angular resolution over the GOODS-North area. In Section 6, we show the effect of the increase in depth and resolution on identifying and studying optical counterparts of sub-mm sources. Here we describe the \hst\ ACS optical observations and the ground-based near-IR observations which are the main focus of this paper.

\subsection{\hst\ imaging}
Deep optical images of the super-map region have been obtained as part of GOODS. The GOODS-North region is approximately 10 arcmin$\times$16.5 arcmin, centred on $12^{\rm{h}}36^{\rm{m}}55^{\rm{s}}, +62^{\circ}14\arcm15\arcs$~\citep{Giavalisco04}. The Advanced Camera for Surveys (ACS) on \hst\ was used to image the region with the F435W, F606W, F775W and F850LP filters (referred to as $B_{\rm{435}}$, $V_{\rm{606}}$, $i_{\rm{775}}$ and $z_{\rm{850}}$, respectively). The images were released to the public in August 2003 and the catalogues followed in December 2003~\citep{Giavalisco04}. 

The depths of this survey are on average a magnitude brighter than the original HDF in $B_{\rm{435}}$, $V_{\rm{606}}$ and $i_{\rm{775}}$~\citep{Williams96}, but over an area that is 32 times larger. The GOODS survey includes the addition of deep $z_{\rm{850}}$-band data (to $z_{\rm{850}}\simeq27$), which extends object selection out to a redshift of $\sim6$ using the Lyman-break technique~\citep{Dickinson04}. Source extractions on the $z_{\rm{850}}$-band images have led to the detection of about 32,000 sources. Photometry has then been carried out in the other ACS bands through matched apertures, providing AB magnitudes in 4 bands for all $z$-detected sources. The measurements given in this paper use the full 5-epoch GOODS ACS images and catalogues~\citep{Giavalisco04}.

\subsection{Near-IR imaging}
As part of the ground-based follow-up to GOODS, near-IR data were obtained using Flamingos at the Kitt Peak National Observatory~\citep[KPNO,][]{Elston03}. These images provide coverage with more uniform sensitivity over the GOODS-North region than the~\cite{Capak04} $HK^\prime$ data, and in the standard $J$ and $K_{\rm{s}}$ filters. The data were reduced by the GOODS team to produce a $K_{\rm{s}}$ selected catalogue with roughly 4,000 sources. These near-IR observations are still ongoing and, due largely to bad weather, the images are currently not as deep as the original goal. Nevertheless, they achieve a depth of $K_{\rm{s}}\simeq22.5$(AB), which is comparable to the near-IR follow-up in other SCUBA surveys (Fox et al.~2002; Smail et al.~2002; Webb et al.~2003). 

The GOODS ACS and near-IR catalogues from SExtractor calculate AB magnitudes through several different size apertures, in addition to providing isophotal and {\tt MAG\_AUTO} photometry measurements~\citep[see][]{BA96}. In this paper, we use matched aperture photometry for determining photometric redshifts and colours, however when single magnitudes are quoted they are the {\tt MAG\_AUTO} values. We quote the magnitude for sources detected at $>5\sigma$ and provide limits, based on the limiting magnitudes at each wavelength, for sources not detected above $5\sigma$.

\section{Deriving optical parameters}

\subsection{Photometric redshifts}

\begin{figure}
\begin{center}
\includegraphics[width=3.0in,angle=0]{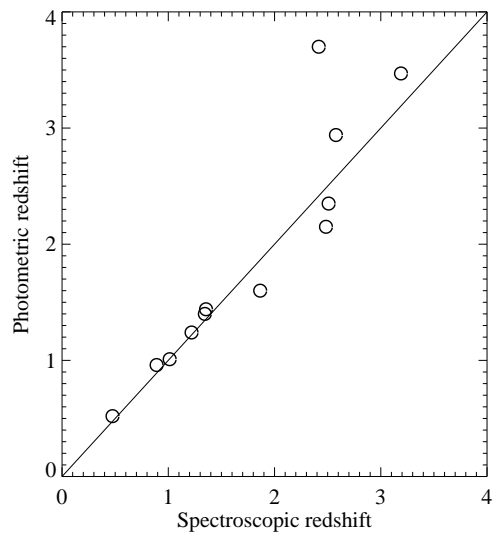}
\caption{Accuracy of photometric redshifts of sub-mm sources in GOODS-North for the 12 sub-mm sources which have spectroscopic redshifts in Cowie et al.~(2004) and Chapman et al.~(2004).
}
\label{fig:specVphot}
\end{center}
\end{figure}

Photometric redshifts have been estimated for a large sample of sources in GOODS-North~\citep{Mobasher04}. For this purpose, available data in the North consist of \emph{U} (KPNO, Capak et al.~2004), \emph{B, V, R, I, z} (SUBARU, Capak et al.~2004), $J$, $K_{\rm{s}}$ (KPNO) and $B_{\rm{435}}$, $V_{\rm{606}}$, $i_{\rm{775}}$, $z_{\rm{850}}$ (ACS \hst), providing up to 12 independent photometric points, although with a high degree of overlap between some of them. The object sample was selected using the ACS $i_{\rm{775}}$-band image and then matched to the ground-based optical and near-IR data. Photometry was carried out through $3\,$arcsec diameter apertures after degrading all images to the Point Spread Function (PSF) of the worst ground-based seeing (FWHM of $1.2\,$arcsec). The final photometric redshift catalogue contains 18,810 sources.

These extensive photometric data have been used to calculate redshifts using both a $\chi^{2}$ minimization technique~\citep{POL82} and a Bayesian method~\citep{Benitez00}. Spectral Energy Distribution (SED) templates consisted of the following galaxy types: E, Sbc, Scd, Im, and starbursts~\citep[see][and references therein]{Benitez00}. Tests of accuracy were performed using the GOODS-South photometric data, which were obtained before the Northern field, and it was found that both statistical techniques were equally consistent with spectroscopic redshifts out to $z\simeq1$~\citep{Mobasher04}. At higher redshifts, the Bayesian method, which considers the redshift probability distribution given the observed colours \emph{and} the overall magnitude, gives better performance. 

We define the ODDS parameter as a quantitative measure of the accuracy of photometric redshifts~\citep{Benitez00}. This corresponds to the integral of the redshift probability distribution within a $0.27(1+z_{\rm{phot}})$ interval centered on $z_{\rm{phot}}$. The ODDS would thus be 0.95 for a Gaussian with width $\sigma=0.067(1+z_{\rm{phot}})$, and is closer to one for well constrained redshifts. In general, we reject any fits that have ODDS$\,<0.90$ and consider the redshift estimate most reliable when ODDS$\,>0.99$~\citep[see][]{Mobasher04}. 

As we discuss in the next section, many of our SCUBA galaxies have optical counterparts identified, while for the others there are several possible counterparts to decide between. We might expect that the sub-mm galaxies will fit best to the starburst or irregular galaxy templates. However, this is based on the assumption that the SEDs of high-redshift galaxies are similar to the SEDs of local ULIRGs, which of course may be incorrect. Moreover, the model templates for the starburst galaxies are not significantly reddened and do not show the ERO colours which are observed in a large fraction of SCUBA galaxies~\citep{Moustakas04}. Therefore, we allow the photometric data for the SCUBA galaxies to be fit to the full range of templates.

In addition to using the ODDS value, we have looked at the redshift probability distributions for each of the sub-mm counterparts. This is useful for verifying a number of things, such as poorly constrained redshifts due to inconsistent photometric points, double peaks in the distribution, and inconsistent redshifts between the $\chi^{2}$ and Bayesian techniques. In those cases, we have been able to fine tune the photometric fitting to provide better estimates for the photometric redshifts of the sub-mm sources. Objects with poor constraints on their photometric redshifts were not included in the analysis of this paper.

Using the catalogue of all spectroscopic redshifts in the HDF-N~\citep[and references therein]{Cohen00,Cowie04,Wirth04}, the photometric redshifts have been shown to be accurate out to $z\simeq2$, with only 5 per cent of sources classified as catastrophic outliers~\citep[see][for results in GOODS-South]{Mobasher04}. At higher redshifts, there are few spectroscopic redshifts with which to compare. Although the accuracy of photometric redshifts might become worse at high redshift, nevertheless galaxies with high photometric redshifts are likely to be at genuinely high redshift. Fig.~\ref{fig:specVphot} shows how consistent the photometric redshifts are for those sub-mm sources that have spectroscopic redshifts in the~\cite{Cowie04} catalogue or the newer Chapman et al.~(2004) paper.

\subsection{CAS structural parameters}
One of the main advantages of the deep ACS images is the high resolution, which allows us to study the structure of the sub-mm counterparts. However, in many cases the optical counterparts are very faint, rendering the determination of detailed morphological information very difficult. In addition to providing a description of the morphology of each source, we have used the `CAS' system~\citep{cc_CAS} to quantitatively describe the optical structure of sub-mm galaxies. This system is appropriate for faint, high-redshift sub-mm sources, as it measures bulk morphological quantities which can be related to a galaxy's past and present physical processes. In the CAS system, three parameters are used to describe galaxies: Concentration (C); Asymmetry (A); and Clumpiness (S). For the optically faint SCUBA sources, the S parameter is unable to give us any reliable information, due to the low resolution of the images (Conselice 2003),
and therefore we focus on C and A. We use the asymmetry index to indicate if the galaxy is undergoing a major merger and the concentration index to tell us how the light is distributed. The asymmetry parameter is basically measured by taking a galaxy, rotating it by $180^{\circ}$, and then subtracting it from the original image (although
there are many details involved in this process, Conselice et al. 2000). The concentration index measures the ratio of the amount of light at the radii that contain 80\% and 20\% of the total amount of light (Bershady et al.
2000). The Petrosian radii can be used to give estimates of the observed sizes of the galaxies~\citep{Petrosian76}. The Petrosian radius is simply the radius where the surface brightness within a thin annulus is a given fraction of the surface brightness within that radius. This fraction is the Petrosian index, $\eta$, and we use 0.2 to define the extent of the galaxy.

It is well known that the morphologies of galaxies evolve both with redshift and wavelength~\citep{cc04}, and therefore we must be very careful in how we interpret the CAS parameters of the sub-mm sample. Most morphology studies are done in the optical, where deep, high resolution imaging is available. However, the longest wavelength ACS filter ($z$) samples the rest-frame near ultraviolet (UV) for sub-mm galaxies at redshifts around 2.5. While morphologies change between the UV and optical for more evolved galaxies, this is less of a problem for starburst galaxies, which look fairly similar at the two wavelengths (Conselice~2004 and references therein). If we want to look at the rest-frame optical morphologies of the sub-mm sample, we need high resolution imaging at infrared wavelengths. A small portion of GOODS-North is covered by deep NICMOS imaging of the HDF~\citep{Dickinson99,Dickinson00}. However, only 5 of the sub-mm sources are within this area. Nevertheless, we have measured the CAS parameters for these sources in both the NICMOS and ACS bands to see how the parameters change as a function of wavelength (see Section 6.3).

\begin{table*} 
\caption{Poisson probabilities for the ACS photometric redshift catalogue used for identification of SCUBA sources. We have calculated statistics for all sources that meet the various colour criteria given in the first column. We give the number density of sources per counterpart search area (i.e. a circle with 7 arcsec radius). $\theta_{5\%}$ is the separation within which there is only a 5 percent probability of finding a source of a given colour by chance. Therefore columns 2 and 3 tell us how rare sources are of each colour criteria, whereas the last 2 columns provide a measure of how likely it is for these sources to be associated with the sub-mm sources at random. We give the number $N$ of the 40 SCUBA sources that have an identification within $7\,$arcsec and then $p\,(\ge N)$ is the random probability of $N$, or more, of the 40 SCUBA objects having at least one counterpart within $7\,$arcsec. We also calculated these statistics using only the $>4\sigma$ sources and found consistent results. $R-K_{\rm{s}}>3.7$ and $i-K_{\rm{s}}>2.5$ correspond to $(R-K)_{\rm{Vega}}>5.3$ and $(I-K)_{\rm{Vega}}>4.0$, respectively. 
}
\label{tab:Pstats}
\begin{tabular}{lllll} 
\hline
Criteria        &  Number per  & $\theta_{5\%}$ & $N$ & $p\,(\ge N)$ \\
                                &  $7\,$arcsec radius &  (arcsec)       & \multicolumn{2}{c}{(for the 40 sources)}  \\\hline\hline
All ACS sources                          &   4.54  &  0.74     & 40  & 0.65 \\
$i-K_{\rm{s}}>2.0$ VRO                   &   0.22    &  3.4      & 14  & 0.017  \\
$R-K_{\rm{s}}>3.7$ ERO                   &   0.05    &  7.4      & 5  & 0.033 \\
$i-K_{\rm{s}}>2.5$ ERO                   &   0.09    &  5.4       & 8  & 0.016 \\
$J-K_{\rm{s}}>1.0$                       &  0.52    &  2.2       & 14  & 0.81 \\
$i-K_{\rm{s}}>2.5$ and $J-K_{\rm{s}}>1.0$ &   0.05    &  7.2       & 7  & 0.0025 \\
\end{tabular}
\end{table*}

\section{Identifying optical counterparts}
Following the methodology for identifying radio and X-ray counterparts developed in Paper II, we take this as our starting point and attempt to make new optical identifications based on what we learned there. We adopt a simple `top-hat' search radius of $7\,$arcsec and consider all sources within this search radius equally. Due to the radio/far-IR correlation~\citep{CY99,CY00}, a radio source within the search radius will be favoured as the counterpart. From the ACS optical images, there can be up to a dozen possible counterparts within our search radius. Using the reliable (ODDS$\,>0.90$, see Section 4.1) photometric redshifts, we can eliminate some of these by imposing the condition that $z>0.5$ for the majority of the sub-mm sources. This condition is determined by the fact that \emph{none} of the 850$\,\mu$m sources are detected at 450$\,\mu$m. Assuming, for example, an Arp220-like SED, we can constrain the redshift from the lack of 450$\,\mu$m flux. The condition $z>0.5$ is quite conservative, given the redshift estimates from the radio/far-IR correlation (see table 5 of Paper II). However, without a large sample of spectroscopic redshifts, we cannot test the accuracy of the radio/far-IR correlation and how it may evolve with redshift. For this reason, we are hesitant to eliminate too many counterparts based solely on the photometric redshift and stick simply to $z_{\rm{phot}}>0.5$. Once these low redshift sources have been removed we then look at the radio-detected (RD) and radio-undetected (RU) sub-populations separately.

\subsection{Radio-detected SCUBA sources}

We have used catalogues from {\it all} the radio observations in GOODS-North (described in Section 3 and Paper II) to identify radio counterparts to our sub-mm sources. We consider radio counterparts within $7\,$arcsec of sub-mm sources as secure and therefore a coincident optical galaxy is a secure counterpart. There are 16 secure RD SCUBA sources in our list and we have found optical counterparts to 15 of these in the GOODS optical and near-IR data. The last one is GN14, or HDF850.1, whose complicated identification is discussed in detail in \cite{Dunlop04}.

GN15, also known as HDF850.2~\citep{Hughes98}, was identified in Paper II with a faint radio source to the North. However, an X-ray detected Lyman-Break Galaxy (LBG) to the west was also mentioned as a possibility. Although there is absolutely no detection of the radio source in any of the optical or near-IR bands, the X-ray/LBG pair is coincident with a red ACS galaxy and has a weak radio signal just below the threshold. This source has a very low 850$\,\mu$m flux and it is likely to be affected by confusion and flux boosting. Therefore although we cannot assign a secure optical counterpart, we now suggest the X-ray/LBG as a tentative identification. 

The closest optical galaxy to the radio counterpart of GN23 is $1.5\,$arcsec away so it is unclear whether this is the same system. The optical photometric redshifts and redshifts estimates from the radio/far-IR correlation (table 5 of Paper II) do not agree, and the colours of the optical galaxy are quite blue. However, there is also a faint smudge in the $K_{\rm{s}}$-band image directly on top of the radio source, below the detection limit, which we identify as the counterpart.
 
One other interesting source worth noting is GN07 which has multi-peaked radio emission within the search radius. The peak of the radio flux is coincident with a very faint ($i_{\rm{775}}\simeq29$) optical source. A fainter radio peak ($<5\sigma$) exists $3\,$arcsec to the west, coincident with a brighter ($i_{\rm{775}}\simeq24$, $z\simeq2$) optical source with bluer colours. In~\cite{Chapman_morph} the sub-mm emission is associated with the brighter optical galaxy, but the colours of this galaxy are not consistent with our other secure optical counterparts at similar redshifts. Sub-mm emission is more likely to trace the radio flux than the optical flux, and therefore we identify the sub-mm emission with the optically faint radio source.

\subsection{Radio-undetected SCUBA sources}

The difference between the RD and RU sub-mm sources is not clear. If we treat them as completely separate populations, then we cannot make any further progress in making optical identifications, and all we would conclude is that each RU SCUBA source has on average half a dozen possible counterparts. An alternative is to use the optical properties of the RD sources to put constraints on the counterparts of the RU sources. In doing this we are not assuming that the properties of the RU must be identical to those of the RD, but we can use their properties as a guide, while considering how the properties might evolve with redshift, magnitude and 850$\,\mu$m flux.

We can use the population \emph{P} probability (see Paper II for a full description) to show how rare it is for the SCUBA sources to have ACS counterparts with particular colours within the $7\,$arcsec search radius. We need to choose magnitude and colour cuts based on what we know about SCUBA galaxies. A lot of effort has been put into understanding the connection between Extremely Red Objects (EROs) and other high redshift populations. EROs are typically defined as having optical-to-infrared colours redder than some threshold, e.g. $(I-K)_{\rm{Vega}}>4.0$, or $(R-K)_{\rm{Vega}}>5.3$~\citep[e.g.][]{Daddi00}. Paper II and other extragalactic SCUBA surveys have found a smaller percentage of EROs coincident with SCUBA galaxies than might have been expected. The $8\,$mJy survey~\citep{Ivison02} find that 19/30 SCUBA sources are radio-detected, 17 of which have an optical counterpart to the depths of the observations, but only 6 of these 17 are identified as EROs. \cite{Webb03} find that 10--20 per cent of the sources in the CUDSS survey are EROs, but state that many more are likely to be EROs below the detection limits. These surveys reached limits of $I\simeq25$ and $K\simeq22.5$. Although the $K_{\rm{s}}$-band data in GOODS North are similar to that of other surveys, the $i_{\rm{775}}$-band data are much deeper, and allow us to detect more EROs. 

\cite{Frayer04} report on the $J-K_{\rm{s}}$ colours of sub-mm galaxies and argue that they can be used to help identify candidate counterparts. They find that the sub-mm galaxies are either faint in the near-IR ($K_{\rm{Vega}}>19$) and very red ($(J-K)_{\rm{Vega}}>3$) or brighter in the near-IR  and with $(J-K)_{\rm{Vega}}\simeq2$. The numbers in Table~\ref{tab:Pstats} tell us that the EROs (as defined by their $i_{\rm{775}}-K_{\rm{s}}$ colours) within $7\,$arcsec of SCUBA sources, which also have red $J-K_{\rm{s}}$ colours, have a $\sim3\times10^{-3}$ chance of being randomly associated. However, the Poisson probability of association for any specific candidate is not low enough for this to be any more than a statistical guide. Nevertheless, Table~\ref{tab:Pstats} shows that $i_{\rm{775}}-K_{\rm{s}}>2.5$ and $J-K_{\rm{s}}>1.0$ together (corresponding to $(I-K)_{\rm{Vega}}>4.0$ and $(J-K)_{\rm{Vega}}>2.0$) is a better overall criterion than simply $i_{\rm{775}}-K_{\rm{s}}>2.5$ or $R-K_{\rm{s}}>3.7$, and that any of these are a vast improvement over selecting ACS sources in a colour-blind way.

The results from other sub-mm surveys, discussed in the previous paragraphs, are biased towards the RD sub-mm sources. So, what do we do to choose among the optical counterparts of the RU sources? We have used two statistical techniques utilizing colours, redshifts and 850$\,\mu$m fluxes to determine the correct optical counterpart for a given SCUBA source. We have applied these to both the RD and RU sub-samples to ensure that we can retrieve the correct counterparts to the RD sources. In other words, we have used the RD sources as a `training-set' to try to find the counterparts of the RU sources.

The first technique uses simple colour cuts, based on the RD colours and fluxes, tuned by hand so that we would select the correct optical identification out of all the candidates within $7\,$arcsec. We also performed a Principal Component Analysis to characterize sub-mm counterparts by the logarithm of their 850 fluxes, their $i_{\rm{775}}-K_{\rm{s}}$ and $J-K_{\rm{s}}$ colours and their $i_{\rm{775}}$ magnitudes. By minimizing the spread in a linear combination of these parameters for the secure identifications, we derive a combination which describes the sample, and then use that and its spread to decide amongst possible counterparts for the RU SCUBA sources. We tried many combinations of parameters and found that the best fit coefficients for the $\rm{log}(S_{850})$, $i_{\rm{775}}-K_{\rm{s}}$, $J-K_{\rm{s}}$ and $i_{\rm{775}}$ magnitude of the sub-mm sources are 1.0, --0.35, 0.30 and --0.008, respectively. Using these values, the most likely optical counterpart within the search radius will have a linear combination of these parameters that is closest to the average value found for the radio-detected sources. Both techniques are successful at finding {\it all} the correct RD counterparts, and using both methods we find tentative counterparts for another 12 of the SCUBA sources. We also tested these techniques to see how many false identifications we might expect. By simulating sub-mm sources at random positions in GOODS-North, our selection techniques found counterparts for these fake sources only 5 per cent of the time. We are therefore confident that these new counterparts are quite trustworthy. This is the first time that plausible counterparts have been found for a significant fraction of radio-undetected sub-mm sources ($12/24$ unique counterparts).

\section{Results and Discussion}

\begin{table*} 
\caption{Properties of optical and near-IR sub-mm counterparts. The SMM ID for the $>4\sigma$ sources is bold-faced. In this table, the RA and DEC are the coordinates of the optical/near-IR counterpart to the sub-mm source. The $i_{\rm{775}}$ magnitude given is the SExtractor {\tt MAG\_AUTO} value~\citep{BA96}, while the photometry for the colours comes from matched apertures. We give the $i_{\rm{775}}$ magnitude to the nearest 0.1, even in cases when the formal error is smaller. If the source is not detected in a band, and we cannot obtain useful constraints from the limits, then the entry is left blank. If the photometric redshift column is blank then the source is too faint to make a reliable (ODDS$>0.90$) photometric redshift estimate. The spectroscopic redshifts are from Cohen et al.~(2000), Cowie et al.~(2004) and Chapman et al.~(2004).
}
\normalsize
\label{tab:colour}
\begin{tabular}{llllllllll}
\hline
SMM ID  &  RA   &  DEC       & $S_{\rm 850\mu m}$ & $i_{\rm{775}}$ mag  &\multicolumn{3}{c}{Optical colours (AB)}            & \multicolumn{2}{c}{Redshift} \\
  & & & (mJy)   & (AB)  &  $R-K_{\rm{s}}$    &    $i_{\rm{775}}-K_{\rm{s}}$        & $J-K_{\rm{s}}$   & Photometric & Spectroscopic \\\hline\hline
\multicolumn{10}{c}{Radio-detected sub-mm sources}\\\hline
\bf{GN04} & 12:36:16.12 & 62:15:14.0 & $5.1\pm1.0$  & $26.2\pm0.1$ &   3.0      & 2.7	& $>1.2$   & 2.94    &  2.578 \\
\bf{GN06} & 12:36:18.40 & 62:15:50.9 & $7.5\pm0.9$  & $27.4\pm0.4$  &   3.6      & 3.3	& $>1.4$   & 1.60 &  1.865 \\ 
\bf{GN07} & 12:36:21.31 & 62:17:08.6 & $8.9\pm1.5$  & $27.8\pm0.4$  &             & 	        &       &     &  \\
\bf{GN11} & 12:36:37.34 & 62:11:51.9	& $7.0\pm0.9$  & $28.1\pm0.3$  &            &            &       &    \\
\bf{GN12} & 12:36:46.07 & 62:14:49.2	& $8.6\pm1.4$  & $26.2\pm0.2$  &   2.2      & 2.0        & $>1.0$   & 1.70   \\   
\bf{GN13} & 12:36:49.72 & 62:13:13.4	& $1.9\pm0.4$  & $21.6\pm0.01$  &   2.0      & 1.6        & 1.0   & 0.52 & 0.475 \\ 
\bf{GN16} & 12:37:00.30 & 62:09:09.9	& $9.0\pm2.1$  & $26.7\pm0.7$ &    $>4.5$      & 4.6        & $>1.6$   & 1.68   &  \\
\bf{GN17} & 12:37:01.60 & 62:11:46.7	& $3.9\pm0.7$  & $27.7\pm0.5$  &    3.6      & 3.2	& 1.6   & 1.72   & $^{\rm{a}}$ \\
\bf{GN19} & 12:37:07.21 & 62:14:08.5	& $10.7\pm2.7$ & $25.4\pm0.1$  & 3.7      & 3.7		& $>1.8$   & 2.15   & 2.484 \\
GN22 & 12:36:06.86 & 62:10:21.7 	& $14.4\pm3.9$ & $24.6\pm0.07$  &   3.5      & 3.2	& 1.4   & 2.35   &  2.509 \\
GN23 & 12:36:08.59 & 62:14:35.8 	& $7.0\pm1.9$   & $>28.0$	 &     & 	&   &    \\
GN25 & 12:36:29.16 & 62:10:46.5	& $4.6\pm1.3$  & 	$22.8\pm0.03$  & 4.1      & 3.1	        & 1.3   & 1.01   & 1.013 \\
GN26 & 12:36:34.53 & 62:12:41.3 	& $3.0\pm0.8$  & $22.7\pm0.02$  &   3.0      & 2.4	& 1.0   & 1.24   & 1.219 \\
GN30 & 12:36:52.77 & 62:13:54.7	& $1.8\pm0.5$  &	$22.7\pm0.01$  & 1.0      & 0.8	        & 0.4   & 1.44   & 1.355 \\
\bf{GN20.2} & 12:37:08.84 & 62:22:02.8 & $11.7\pm2.2$ & $24.7\pm0.07$  &   &    &    &  3.91  &   \\\hline
\multicolumn{10}{c}{Radio-undetected sub-mm sources} \\\hline 
\bf{GN01} & 12:36:06.70 & 62:15:51.0    & $7.3\pm1.5$   & $23.3\pm0.02$   & 1.2    & 1.1   & 0.7 &   3.70   &   2.415   \\
\bf{GN03} & 12:36:09.12 & 62:12:54.4	& $16.8\pm4.0$  & $27.6\pm0.6$  &       &         &       & 2.10   \\
\bf{GN05} & 12:36:19.13 & 62:10:04.4	& $6.7\pm1.6$   & $24.9\pm0.2$   & 4.0  & 2.8     & $>1.3$   & 2.60    \\ 
\bf{GN08} & 12:36:21.38 & 62:12:53.2	& $12.5\pm2.7$  & $24.0\pm0.05$  & 1.8    & 1.5    & $>1.1$      & 2.12     \\
\bf{GN10} & 12:36:33.25 & 62:14:11.5	& $11.3\pm1.6$  & $23.7\pm0.02$  & 1.3    & 1.2	& 0.4     & 1.44 &  1.344  \\
\bf{GN15} & 12:36:55.79 & 62:12:01.1	& $3.7\pm0.4$   & $24.3\pm0.07$  & 2.8    & 2.2	& 0.9	    &  & 2.743  \\
\bf{GN20} & 12:37:11.69 & 62:22:13.5  & $20.3\pm2.1$  & $26.5\pm0.2$  &        &          &           &       \\
GN24 & 12:36:12.01 & 62:12:22.2	& $13.7\pm3.6$  & $24.7\pm0.09$   & 2.5  & 2.2      & $>1.4$   & 2.91      \\
GN31 & 12:36:53.59 & 62:11:15.6	& $2.8\pm0.8$   & $23.0\pm0.03$   & 1.7    & 1.2	& 0.7	    & 0.96   & 0.890 \\ 
GN32 & 12:36:58.72 & 62:14:59.3	& $5.3\pm1.4$   & $27.8\pm0.4$   &        & 	&    	    &    \\ 
GN34 & 12:37:07.28 & 62:21:15.8	& $5.6\pm1.6$   & $23.9\pm0.04$  & 4.0   &  2.9   & 1.1   &  1.00  &   \\
GN37 & 12:37:38.30 & 62:17:36.5  & $6.8\pm1.9$   & $23.1\pm0.02$   & 0.9    & 0.9    & 0.8  &  3.47  &   3.190 \\\hline
\normalsize
\end{tabular}
\medskip
\\
$^{\rm{a}}$\,Cohen et al.~(2000) list this galaxy at $z=0.884$, where the source of this spectroscopic redshift is the Hawaii group. However, Cowie et al.~(2004) have tabulated all previously published redshifts and they do not list this particular redshift. Due to this inconsistency, we have left this spectroscopic redshift out of our analysis.\\
\end{table*}

Our sample contains 39 sub-mm sources with optical and near-IR imaging in GOODS-North. We have identified unique optical counterparts for 15 RD and 12 RU sub-mm sources, which makes 72 per cent of our sample optically identified. An additional $\sim18$ percent have multiple counterparts that meet our selection criteria. Therefore, about 90 per cent of the sub-mm sources in GOODS-North have optical counterparts, making it close to complete. These results are similar if we only consider the $>4\sigma$ sub-mm sources or the brighter sub-mm sources. The ACS and near-IR Flamingos images have revealed new optical counterparts, not present in shallower surveys. Fig.~\ref{fig:postage} shows 10 arcsec ACS, or near-IR images, of the sub-mm counterparts. Table~\ref{tab:colour} lists the sub-mm sources which have optical counterparts, along with their associated colours and redshift information. Note that we have left out GN14, also known as HDF850.1, since the counterpart to this source is known to be complicated, as discussed in detail in~\cite{Dunlop04}. The RD counterparts are all considered `secure', while the RU counterparts are marked as `tentative'. The SMA and ALMA will be able to resolve the exact sub-mm positions and make secure identifications in the absence of a radio counterpart. In addition, $Spitzer$, specifically MIPS 24$\,\mu$m, will soon improve the situation. Although the resolution is poor, the MIPS 24$\,\mu$m population should show a high correlation with both the RD and RU SCUBA sources. Meanwhile, we can make progress by studying the properties of these new RU counterparts and assessing how they compare to the RD sources, as well as comparing with what we might have expected for sub-mm galaxies.

\subsection{Redshift distribution} 

\begin{figure*}
\begin{center}
\includegraphics[width=6.0in,angle=0]{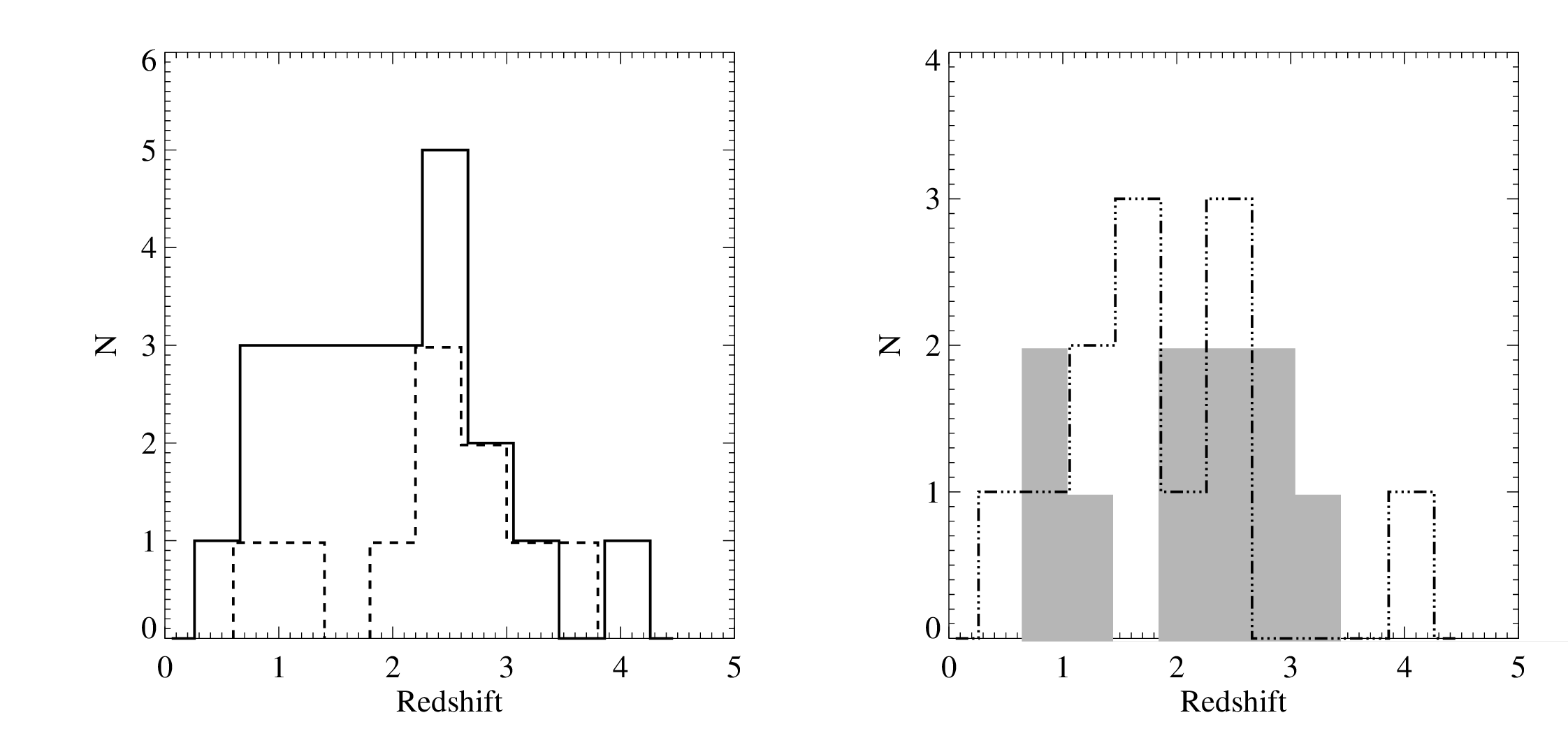}
\caption{Redshift distribution of sub-mm sources. The plot on the left shows the distribution for {\it all} of our sub-mm sources (solid line), where we have plotted the spectroscopic redshift when available and the photometric redshift otherwise. The dashed line is the distribution from Chapman et al. (2003a) of radio-selected sub-mm-bright sources. On the right, the distribution of our sub-mm sample has been split up into the radio-detected sources (dot-dash line) and the radio-undeteted sources (shaded region).
}
\label{fig:photzdist}
\end{center}
\end{figure*}

\begin{table} 
\caption{Median redshifts (photometric unless spectroscopic is available) for sub-mm galaxies in GOODS-North. 
}
\label{tab:medianz}
\begin{tabular}{llll} 
\hline
Source type &  No.~of sources  &  Median redshift & Quartile range\\\hline
\multicolumn{4}{c}{Bright sub-mm sources with $S_{850}>5\,$mJy} \\\hline 
Radio-detected      & 7      & 2.48 & 1.70--2.58       \\
Radio-undetected    & 8     & 2.27  & 1.72--2.76      \\
All                 & 15      & 2.42 & 1.70--2.60     \\\hline 
\multicolumn{4}{c}{All sub-mm sources} \\\hline 
Radio-detected      & 12       & 1.71 & 1.29--2.50     \\
Radio-undetected    & 10       & 2.27 & 1.34--2.74      \\
All                 & 22       & 1.98 & 1.34--2.58     \\
\end{tabular}
\end{table}

The photometric redshifts for our sub-mm sample are listed in  Table~\ref{tab:colour}. We have only given the photometric redshifts when it has an ODDS parameter of $>0.90$. 13 of the counterparts have published spectroscopic redshifts, and in all but one case the photometric redshift agrees to within 15 per cent (see Fig.~\ref{fig:specVphot}). Although this has been indicated in some other studies~\citep[e.g.][]{Clements04}, this new comparison effectively demonstrates that optical/near-IR photometric redshifts can provide a good estimate for the redshift of sub-mm galaxies in the absence of spectroscopy. For the analysis of this paper, we have adopted the spectroscopic values for these sources. Redshift constraints, based on the radio/far-IR correlation (Carilli \& Yun~1999,~2000) have been discussed in Paper II and are in general agreement with the optical photometric redshifts, within the error bars. In Paper II, one disparity was the photometric redshift of GN22 from~\cite{Barger02} which was inconsistent with the radio photometric redshift. This source now has a secure ACS photometric redshift that is within the radio photometric redshift range.

\begin{figure*}
\begin{center}
\includegraphics[width=6.0in,angle=0]{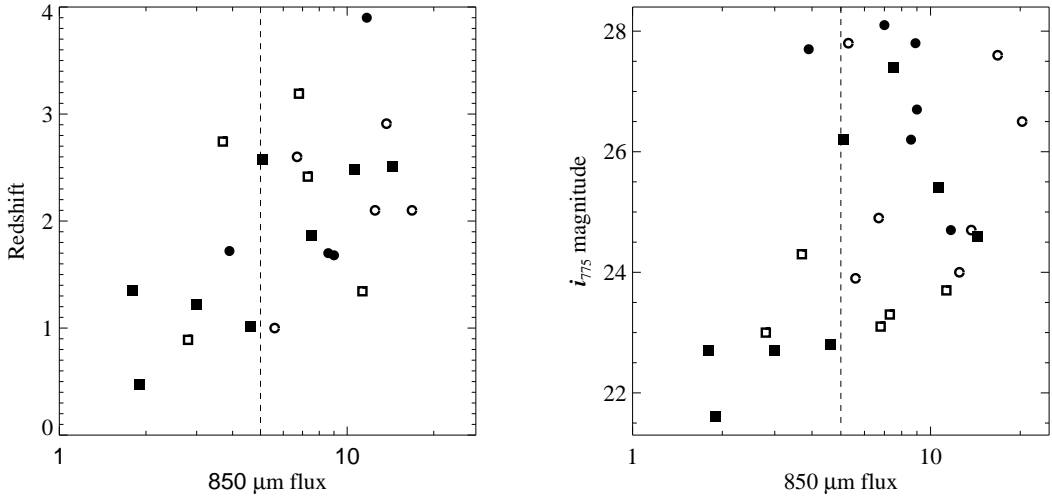}
\caption{Redshift and $i_{\rm{775}}$ magnitude as a function of 850$\,\mu$m flux density on a logarithmic scale. The solid symbols are the RD sources while the open symbols are the RU sources. The squares denote sources which have spectroscopic redshifts, while the circles have photometric redshift estimates. The dashed horizontal line indicates a flux of $5\,$mJy, above which the majority of the Chapman et al. (2003a) sources lie. The lack of sources in the top left quadrant of both panels is intriguing; it could be due to a number of effects, including evolution.
}
\label{fig:F850}
\end{center}
\end{figure*}

In the left panel of Fig.~\ref{fig:photzdist}, we plot the distribution of redshifts for {\it all} our sub-mm sources, which has a median redshift of 2.0 and a quartile range of 1.3--2.6. For comparison, we have also plotted the distribution from spectroscopic studies. \cite{Chapman_nature} have targetted a sample of sub-mm bright ($S_{\rm 850\,\mu m}\gsim5$) RD sources to obtain a median redshift of 2.4. Our sample has a significant number of fainter 850$\,\mu$m sources, which tend to be at lower redshifts (Fig.~\ref{fig:F850}). Table~\ref{tab:medianz} shows that we find a median of 2.5 if we are restricted only to this brighter sub-population. A Kolmogorov-Smirnov (KS) test does not find that our full distribution, or the bright RD sub-sample, is significantly different from that of \cite{Chapman_nature}. Our median value for all sub-mm sources is slightly lower than the median from spectroscopic redshifts only, which may be because photometric redshift estimates are not affected by the `redshift-desert' at $z\simeq1.5$ apparent in the \cite{Chapman_nature} distribution. One might expect our full distribution to go out to higher redshifts since we are not constrained by the radio observations. Fig.~\ref{fig:photzdist} shows no evidence for such a high redshift tail. However, we caution that our results may be biased by selection effects (i.e. galaxies must be detected in the optical to be in our sample) and systematic uncertainties in the photometric redshifts.

The right panel of Fig.~\ref{fig:photzdist} shows how the distributions of our RD (dot-dash line) and RU (shaded region) sources compare. While the former clearly peaks around $z\simeq2$, the RU sources are spread over a broad range of redshifts. However, with a KS test, we cannot rule out the possibility that these samples are drawn from the same distribution. Assuming a ULIRG SED, we generally expect that SCUBA sources at redshifts $\le2$ to have radio counterparts, although it is possible for the lower sub-mm flux sources to be below the radio detection limit even at lower redshifts. The 2 RU sources with the lowest 850$\,\mu$m flux also have low redshifts which is consistent with this idea. The median redshifts for these different sub-samples are shown in Table~\ref{tab:medianz}. For the full sub-mm sample, the median for the RU sources is slightly higher, however the quartile ranges overlap significantly, and the low number statistics make it difficult to draw any firm conclusions. 

Our current understanding of sub-mm sources comes primarily from the RD sub-population, therefore determining the differences between RD and RU sub-mm sources is very important. It is entirely possible that the RU SCUBA sources sample into the high-redshift tail of the distribution. On the other hand, as temperature and redshift are degenerate, the lack of radio flux could also be because these galaxies are cooler. It may be that it is not completely correct to assume that the SEDs of high-redshift SCUBA galaxies are similar to those of local ULIRGs, in which case the RU SCUBA sources could be intrinsically different from the radio-detected population. Although we find little difference in their optical properties (see Section 6.2), the information is still rather limited, and this is another area where \emph{Spitzer} data will help.

\begin{figure*}
\begin{center}
\includegraphics[width=6.0in,angle=0]{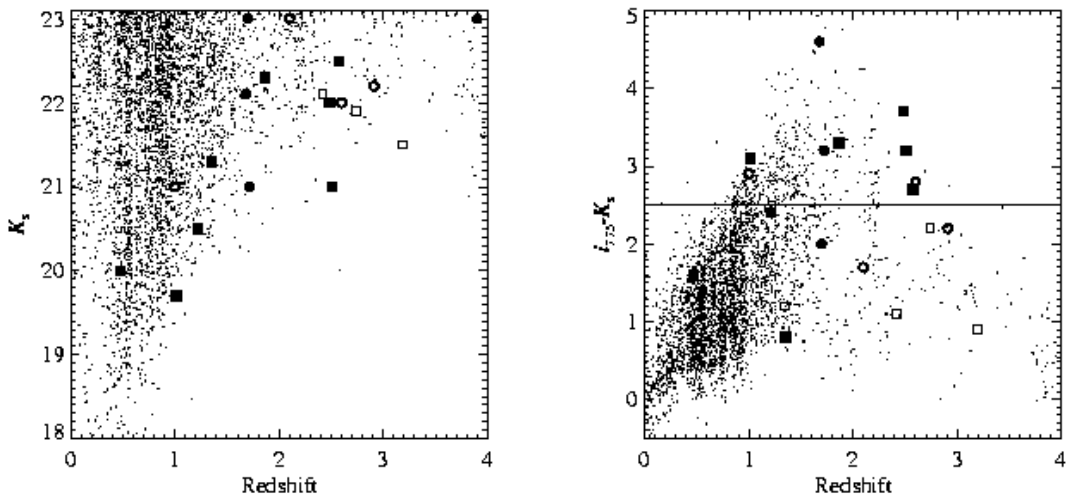}
\caption{Magnitudes and colours of sub-mm counterparts as a function of redshift, compared with field galaxies. All galaxies from GOODS-North with reliable photometric redshifts (ODDS$\,<0.90$) and $5\sigma$ detections in $K_{\rm{s}}$ (left and right panels) and $i_{\rm{775}}$ (right panel) are plotted along with the sub-mm galaxies. Again, the solid symbols are the RD sources while the open symbols are the RU sources, with squares and circles denoting sources with spectroscopic and photometric redshifts, respectively. Although there are not as many sample galaxies to compare with at higher redshifts, we see that the sub-mm counterparts tend to be found at redder colours and higher redshifts than the bulk of galaxies. The horizontal line is the conventional ERO cut-off.
}
\label{fig:photVcolour}
\end{center}
\end{figure*} 

Another thing we can examine is variations in the optical properties with SCUBA brightness. In Fig.~\ref{fig:F850} we plot the 850$\,\mu$m flux as a function of redshift and $i_{\rm{775}}$ magnitude. Although the dynamic range in the sub-mm flux density is small, we do notice that the fainter sub-mm sources stand out from the rest of the sources in both plots (a similar result was tentatively suggested earlier by Ivison et al.~2002). Recall that because of the negative K-correction at 850$\,\mu$m, the observed flux density for a galaxy with a specific luminosity is essentially constant past $z\sim1$, meaning that fainter SCUBA sources are typically intrinsically less luminous~\citep{Blain02}. 

It is not clear that there is one reason why there are trends in Fig.~\ref{fig:F850}, however we suspect it is a mixture of several effects: 

(i) A higher 850$\,\mu$m flux should correspond to a higher far-IR luminosity which implies a higher dust content, implying that it is more likely for brighter sub-mm sources to be fainter in the optical.

(ii) Going to higher redshifts we are looking at more volume, therefore the most luminous and rare far-infrared sources are more likely to be at high redshift. This is true even if we do not include the effects of evolution.
 
(iii) The evolution of ULIRGs is a strong function of redshift~\citep[][and references therein]{Blain02}, so the highest far-IR luminosity sources (and hence higher 850$\,\mu$m flux sources) are more common at high redshift. The fainter galaxies at lower redshifts could be less extreme star-forming galaxies, consistent with LIRGs, rather than ULIRGs, for example~\citep{Sajina03}. 

(iv) There are possible selection effects to do with counterpart selection, limiting magnitudes of the optical data and the sources for which we can get reliable photometric redshifts. 

(v) We have small number statistics, and a small dynamic range in the 850$\,\mu$m flux density.
 
We tested the correlations in Fig.~\ref{fig:F850} by choosing an optical counterpart at random within the sub-mm search radius and plotting the results. These plots did not show any of the correlations that are found for the true sub-mm counterparts. We therefore conclude that these are genuine effects in our data, although we have no clear explanation as yet. 

\subsection{Optical and near-IR colours}   

Table~\ref{tab:colour} lists the colours and $i_{\rm{775}}$-band magnitudes for all the sub-mm counterparts. The average magnitude in $i_{\rm{775}}$ is around 25 for both the RD and RU sub-mm sources. All but one of the counterparts is detected in $i_{\rm{775}}$, indicating that the depth of this survey is sufficient for identifying the majority of sub-mm sources in the optical. The left panel of Fig.~\ref{fig:photVcolour} shows that a significant number of the sub-mm counterparts track the bright envelope of the Hubble diagram, i.e. they are among the most luminous galaxies at their redshifts. The remainder are within 1--2 mags of the bright envelope of the luminosity distribution. The bright $K_{\rm{s}}$ magnitudes make the sub-mm sources much redder than other populations at similar redshifts (right panel of Fig.~\ref{fig:photVcolour}). The average $i_{\rm{775}}-K_{\rm{s}}$ colour for all the ACS sources in GOODS-North is 1.4 with a standard deviation of 0.8. This is equivalent to $(I-K)_{\rm{Vega}}=2.9$. If we restrict ourselves to high redshift ($z>1$), then the average is higher ($i_{\rm{775}}-K_{\rm{s}}\simeq2.0$, standard deviation of 0.9), however the sub-mm sources are redder still, with an average colour of $i_{\rm{775}}-K_{\rm{s}}\simeq2.3$ (standard deviation of 1). 

Fig.~\ref{fig:photVcolour} does not show a striking difference between the RD and RU sources, however the RU sources are slightly less red at a given redshift. An ERO ($(I-K)_{\rm{Vega}}>4.0$) as defined through these filters corresponds to $(i_{\rm{775}}-K_{\rm{s}})_{\rm{AB}}>2.5$. With this criterion, 65 per cent of the RD sources with a near-IR counterpart are EROs, while only 22 per cent of the RU sources with a counterpart in the near-IR are EROs. This is a much higher ERO fraction for the RD sources than previous SCUBA surveys, which is because of the depth of the ACS data. If our detection limit in $i_{\rm{775}}$ had only been 25 (which is typical for much of the optical follow-up to SCUBA surveys), then we would only have found 2 EROs as counterparts to RD sources. Furthermore, \emph{half of our optical counterparts would not have been detected at this lower magnitude limit}. Therefore the depth in $i_{\rm{775}}$ is key for determining the degree of redness and finding counterparts.

$J-K$ colour can also be used to give a measure of the redness of a galaxy, particularly at higher redshifts. \cite{Franx03} argue that high redshift galaxies ($z>2$) can be selected based on $(J-K)_{\rm{Vega}}>2.3$, which corresponds to $(J-K_{\rm{s}})_{\rm{AB}}>1.3$. The sub-mm sources that make this near-IR colour cut are mostly around redshift 2, and there are higher redshift sources with lower $J-K_{\rm{s}}$ values. Since our near-IR data do not currently reach the same depth as the HDF-South near-IR data used in \cite{Franx03}, we cannot detect as many sources with extreme near-IR colours. However, several of our sources, while detected in $K$, are below the detection limit in $J$. We do not detect any sub-mm sources with $(J-K)_{\rm{Vega}}>3$ and therefore cannot test the idea in~\cite{Frayer04}, that sub-mm galaxies, which are near-IR faint, are extremely red in the near-IR. However, the $J-K_{\rm{s}}$ colours we do measure are consistent with the colours found in~\cite{Frayer04} for the near-IR bright sub-mm galaxies. As we saw in Fig.~\ref{fig:photVcolour}, our sub-mm sources are bright in the near-IR at all redshifts.

\begin{figure}
\begin{center}
\includegraphics[width=3.0in,angle=0]{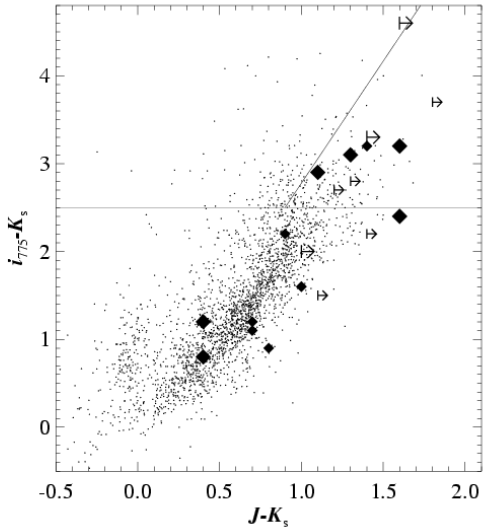}
\caption{Near-IR colour-colour plot. Dots are field galaxies from the GOODS catalogue. Sub-mm sources with reliable photometry through all 3 filters are presented as diamonds, with limits shown as arrows. The large symbols are the sub-mm sources with $1<z<2$. According to Pozzetti \& Mannucci (2000), within this redshift range, sources which are starburst EROs, as opposed to ellipticals at $z\simeq1$, are expected to have red $J-K_{\rm{s}}$ colours and fall to the right of the diagonal line in the colour-colour plot. It is a good check to see that all of the sub-mm sources which are EROs and are at $1<z<2$ are consistent with starburst galaxies. The horizontal line is the the traditional ERO cut-off through these filters.
}
\label{fig:colourcolour}
\end{center}
\end{figure}

\cite{PM00} discuss the two main types of EROs, namely elliptical galaxies at $z\sim1$ and dusty starburst galaxies. When plotted on a near-IR colour-colour diagram, the EROs with $1<z<2$ are clearly a separate population. \cite{BW04} show that outside this redshift range ellipticals contaminate the starburst section of the plot and the near-IR colours no longer separate the two ERO populations. In Fig.~\ref{fig:colourcolour}, we show the near-IR colour-colour plot for the ACS field galaxies with the sub-mm galaxies highlighted. Above the ERO cutoff, all the sub-mm galaxies lie on the starburst side of the \cite{PM00} cut-off. Again, the RU sub-mm sources are similar to the RD sources.

While the differences are not huge, it is interesting that the RU sub-mm sources show less extreme colours than the RD sources, even though they were effectively selected with the same colour cuts. This is also seen in a lower fraction of EROs found for the RU sources. We might expect the RU sources to be at higher redshifts and also to be redder, but we have shown in the previous section that the RU and RD sources appear to have quite similar redshift distributions. This difference seems to be significant and not obviously due to selection effects.


\subsection{Internal structure}  

\begin{figure*}
\begin{center}
\includegraphics[width=6.0in,angle=0]{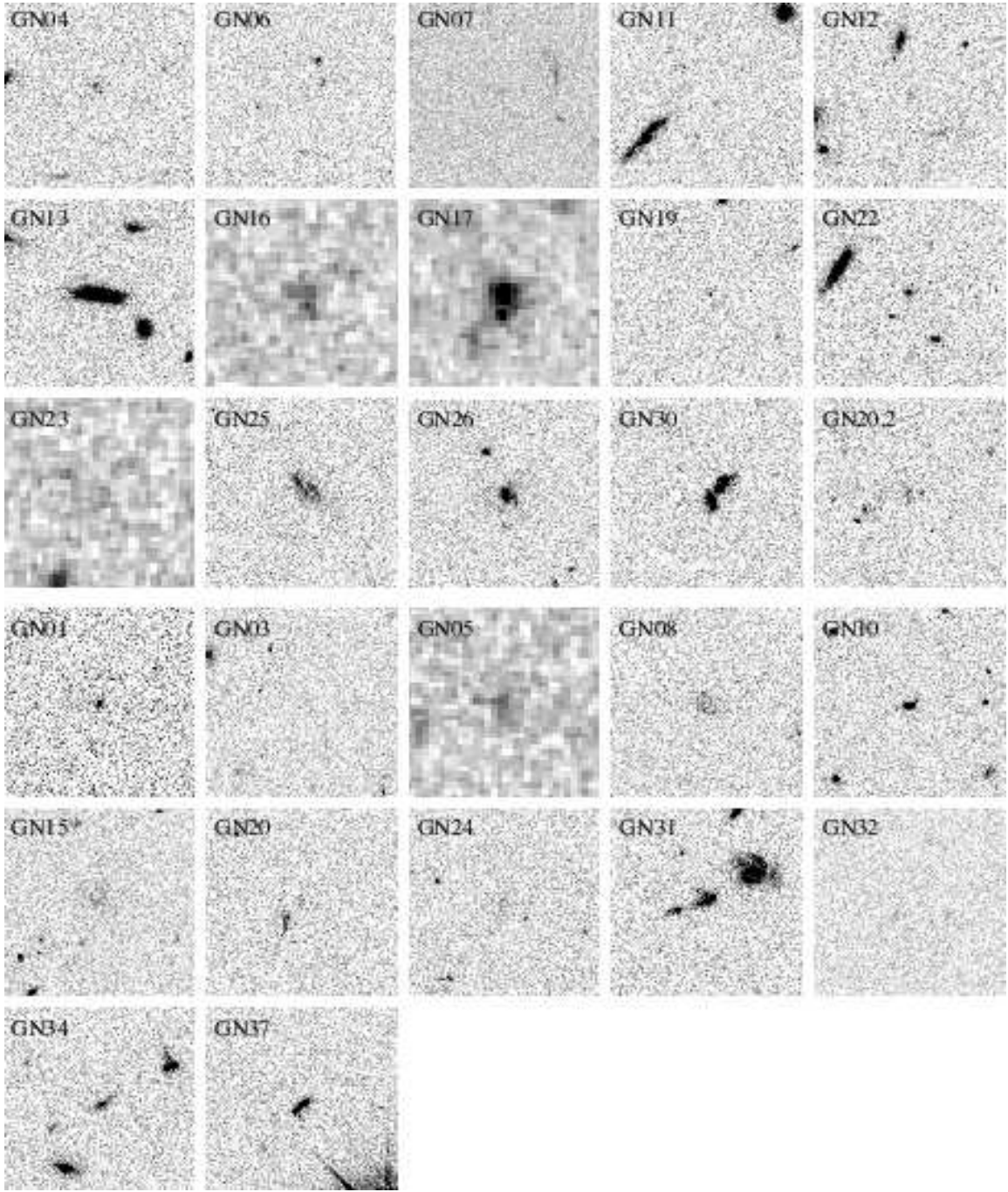}
\caption{Optical or near-IR images of sub-mm counterparts. Images are all 10 arcsec across and are centred on the optical counterpart. At this scale and contrast level, the optical counterpart is sometimes very faint, however these images provide `finding charts' for identifying the position of the optical counterpart. The images are from the ACS $i_{\rm{775}}$-band except for GN32 which is an ACS $z_{\rm{850}}$-band image and GN16, GN17, GN23 and GN05 which are Flamingos $K_{\rm{s}}$ images. The first three rows are the radio-detected sources and the last three are the radio-undetected sources. North is up and east is to the left in these images.
}
\label{fig:postage}
\end{center}
\end{figure*}

\begin{figure*}
\begin{center}
\includegraphics[width=6.0in,angle=0]{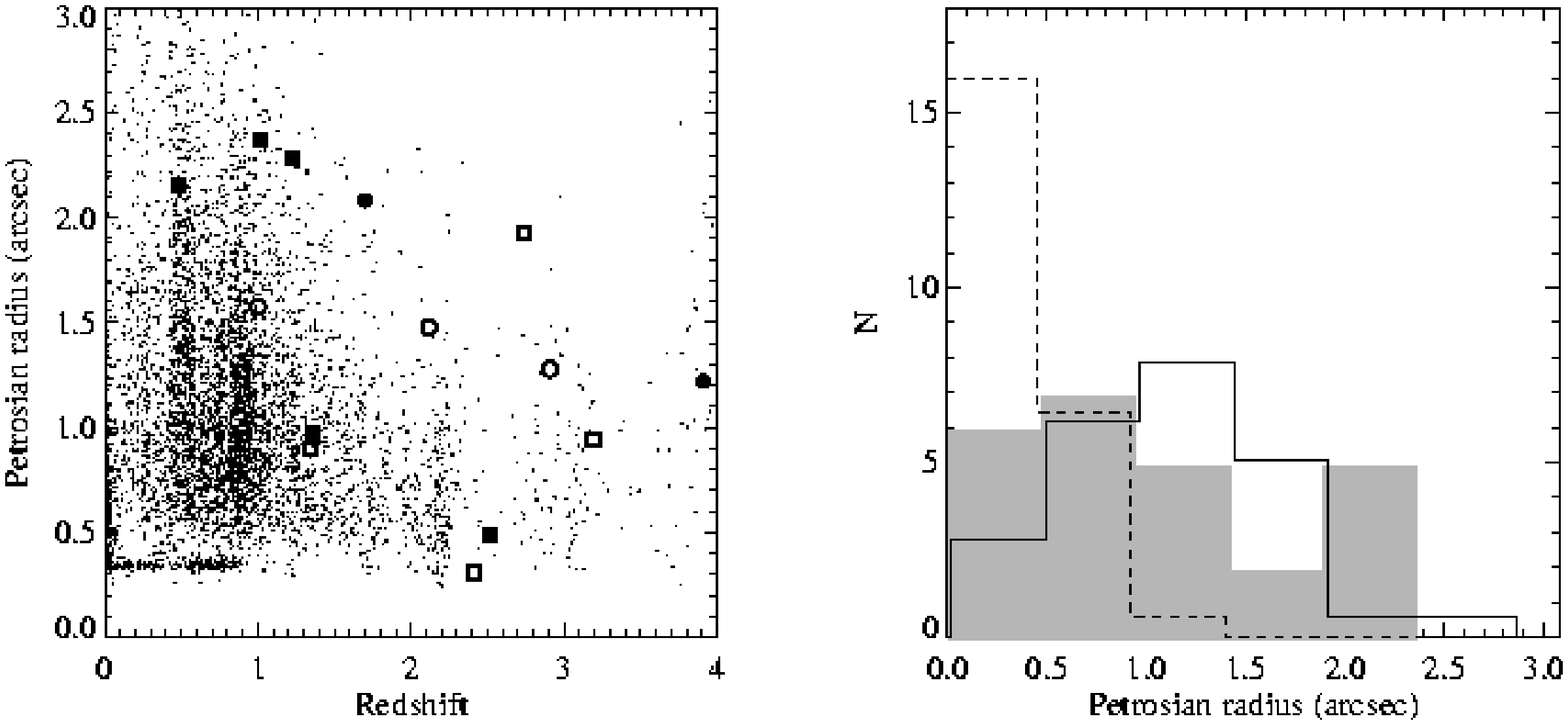}
\caption{Sizes of SCUBA galaxies. The left panel shows the Petrosian radius (in arcsec) as a function of redshift. We have plotted all sub-mm sources with $i_{\rm{775}}<25$ and a reliable photometric redshift (ODDS$>0.90$). Note that we have restricted the magnitude since it becomes difficult to measure the morphology parameters at fainter magnitudes. Therefore we are comparing the brighter sub-mm counterparts to the brigher field galaxies at each redshift. Solid symbols are the RD sub-mm sources and the open symbols are the RU sub-mm sources. Sources with spectroscopic redshifts are plotted as squares and those with photometric redshifts are denoted by circles. Note that there is an excess of galaxies with a measured Petrosian radius of $\simeq0.3$ as this is the minimum measurable radius. The right panel shows the distribution of Petrosian radius. The shaded region is the distribution of our sub-mm sample. The solid and dashed lines are samples of local ULIRGs and normal galaxies, respectively, redshifted to $z\simeq2$. These distributions have been normalized to the same total number and slightly offset for clarity. In terms of size, the sub-mm sources are clearly not normal galaxies and appear similar to the redshifted ULIRGs.
}
\label{fig:petr}
\end{center}
\end{figure*}

Sub-mm galaxies at high redshift are expected to evolve into the massive ellipticals in the local Universe~\citep[e.g.][]{Lilly99}. Their luminosities imply high star-formation rates and their volume density is consistent with local giant elliptical galaxies~\citep[e.g.][]{Chapman_nature}. To investigate if sub-mm galaxies are indeed evolving into massive ellipticals, we can study their morphologies. The few sub-mm galaxies that have been studied with deep, high-resolution optical observations show a range of morphologies, with a significant number showing asymmetries, consistent with early-stage mergers (Smail et al.~1998; Ivison et al.~2002; Chapman et al.~2003b; Conselice et al. 2003; Clements et al.~2004).

\begin{table*}  
\caption{Morphological parameters for sub-mm sources. `C' and `A' are the concentration and asymmetry parameters, respectively, while `dC' and `dA' are their associated errors (see Conselice 2003). `Petr R' is the Petrosian radius in units of arcsec. This table gives the raw CAS parameters, which have not been corrected for redshift. Bold-faced redshifts are spectroscopic (Cowie et al.~2004, Chapman et al.~2004), while the rest are photometric. The Comments column is based on a visual inspection of the ACS images. LSB is low surface brightness.
}
\normalsize
\label{tab:CAS}
\begin{tabular}{lllllllll}
\hline
SMM ID    &  $i_{\rm{775}}$ mag     & Redshift & C & dC & A & dA & Petr R (\arcsec) & Comment  \\\hline\hline

\multicolumn{9}{c}{I. Low redshift, normal CA parameters: $z<1.5$, $\rm{A}<0.15$} \\\hline 
GN13	&  21.6 & \bf{0.475} & 2.98  & 0.05  & 0.12  & 0.03  & 2.15    & normal, disk-like        \\
GN34	&  23.9 & 1.00        & 3.25  & 0.06  & 0.09  & 0.13  & 1.57 &  large, disk-like  \\
GN25	&  22.8 & \bf{1.013} & 2.75  & 0.03  & 0.06  & 0.09  & 2.37    & asymmetric         \\

\hline
\multicolumn{9}{c}{II. Low redshift, higher asymmetry: $z<1.5$, $\rm{A}>0.15$} \\\hline
GN31	&  23.0 &  \bf{0.890}       & 2.13  & 0.06  & 0.17  & 0.06  & 1.26   & diffuse , LSB         \\ 
GN26	&  22.7 & \bf{1.219} & 3.43  & 0.04  & 0.28  & 0.05  & 2.29    & asymmetric         \\
GN30	&  22.7 & \bf{1.355} & 3.06  & 0.10  & 0.36  & 0.02  & 0.97   & double, asymmetric, bright \\
GN10	&  23.7 & \bf{1.344} & 3.45  & 0.12  & 0.30  & 0.03  & 0.90    & compact            \\

\hline
\multicolumn{9}{c}{III. Mid/high redshift, negative/noise-dominated asymmetry $z>1.5$, $\rm{A}<0$ or $\rm{dA}\gsim0.1$} \\\hline 
GN06	&  27.4  &  \bf{1.865}       & 2.92  & 0.07  & --0.03  & 0.17  & 0.89   &  faint smudge         \\ 
GN12  	&  26.2  & 1.70        & 3.32  & 0.10  & --0.04  & 0.15  & 0.65  &  faint        \\ 
GN17	&  27.7 & 1.72        & 1.99  & 0.03  & 0.03  & 0.18  & 2.08   & faint, fuzzy           \\
GN08	&  24.0 &  2.12       & 2.23  & 0.04  & 0.05  & 0.13  & 1.48   & faint, large, diffuse    \\
GN19	&  25.4 & \bf{2.484}  & 3.05  & 0.09  & 0.03  & 0.13  & 0.80  & compact, some/diffuse    \\
GN22    &  24.6  & \bf{2.509} & 2.91  & 0.17  & --0.12  & 0.08  & 0.49    & normal, compact          \\
GN04	&  26.2 & \bf{2.578}  & 2.91 & 0.17  & 0.11  & 0.09 & 0.44 & pair, double\\
GN15	&  24.3  & \bf{2.743} & 2.22  & 0.04  & --0.15  & 0.20  & 1.92   & asymmetric structure       \\
GN24	&  24.7 & 2.91        & 2.69  & 0.04  & 0.04  & 0.15  & 1.28  & very diffuse, LSB            \\
GN20.2  &  24.7  & 3.91       & 2.40  & 0.06  & --0.12  & 0.25  & 1.22  &  diffuse, LSB   \\

\hline
\multicolumn{9}{c}{IV. High redshift, higher asymmetry: $z>2.4$, $\rm{A}>0.15$} \\\hline 
GN01    &  23.3 & \bf{2.415} & 2.79  & 0.28  & 0.23  & 0.003  & 0.31   & compact            \\
GN37    &  23.1 &  \bf{3.190} & 2.41  & 0.08  & 0.43  & 0.03  & 0.94  & double, asymmetric     \\

\hline
\multicolumn{9}{c}{V. Unknown redshift} \\\hline 
GN07	&  27.8  &           & 2.61  & 0.17  & --0.70  & 0.30  & 0.43  &  very faint         \\
GN11	&  28.1 &            & 2.28  & 0.30  & 0.34  & 0.14  & 0.36   &  compact        \\
GN20    &  26.5 &            & 2.37  & 0.17  & 0.14  & 0.19  & 0.42  &  faint, asymmetric          \\
GN32	&  27.8 &            & 2.57  & 0.20  & 0.11  & 0.15  & 0.43    & very faint           \\ 
\hline

\normalsize
\end{tabular}

\end{table*}

Fig.~\ref{fig:postage} gives some indication of morphology for our sample. Some of the brighter galaxies are clearly undergoing mergers, although most are too faint to tell much from just visual inspection. The effects of distance and extinction on the morphologies of the sub-mm sample are very significant; most of the counterparts are very faint and therefore the CAS parameters are difficult to measure accurately. While in the local Universe we define a merger as having an asymmetry value of $>0.35$~\citep{cc_submm}, we cannot make such a cut at higher redshifts, since the asymmetry parameter decreases as a function of wavelength, as it gets harder to measure. We must therefore use the redshifts (in most cases photometric) in order to compare the structure of sub-mm sources to the morphologies of other high-redshift sources within GOODS North. We can do this as a function of $i_{\rm{775}}$ magnitude as well, although we may be biasing the comparison sample by only choosing other faint galaxies. 

Table~\ref{tab:CAS} summarizes the morphologies of the sub-mm sample in GOODS-North. In addition to measuring the CAS parameters and the radius, we have examined the images and classified each by eye (see Fig.~\ref{fig:postage}). The results of this classification are listed in the comments column of the Table. Even when looked at as a function of redshift, the sub-mm galaxies show a wide range of morphologies. Although the concentration, asymmetry and radius on their own do not separate the sub-mm sources from the field galaxies, when coupled with the redshift and $i_{\rm{775}}$ magnitude, we are able to classify our sample into several groups, namely: low redshift, normal CA parameters (Group I); low redshift, higher asymmetry (Group II); mid/high redshift, negative/noise-dominated asymmetry (Group III); high redshift, higher asymmetry (Group IV); and unknown redshift (Group V). Groups II and IV both show high asymmetry, which is consistent with merging systems. Unfortunately the largest group is III, which contains either negative asymmetries, meaning that the sky dominated the calculation, or poorly constrained asymmetries. In an attempt to remove the strong effects of the sky, we tried measuring the parameters at half the radius, but found no improvement and mildly reduced asymmetries. This is expected since the asymmetry parameter tends to pick up the larger scale structure in the galaxies which may be removed at smaller radii. All parameters in Table~\ref{tab:CAS} were measured in the ACS $z_{\rm{850}}$-band image because we want to sample as close as we can to the rest-frame optical. However, we also looked at the parameters in the $i_{\rm{775}}$ and $i_{\rm{775}}+z_{\rm{850}}$ image. While the latter image has better signal-to-noise ratios, it did not make a significant improvement to the measurements. 

Fig.~\ref{fig:petr} shows how the sizes of sub-mm galaxies compare to other high-redshift galaxies. In the left panel, we see that the Petrosian radius of the sub-mm galaxies is larger than for ACS field galaxies at essentially all redshifts. A KS test tells us that for $z<2$ or $i_{\rm{775}}<25$ (which is more or less equivalent) the probability that the SCUBA galaxy sizes are drawn from the size distribution of field galaxies is less than 5 per cent. At higher redshifts or fainter magnitudes, the hypothesis that they have the same size distribution cannot be rejected because of smaller numbers. In terms of asymmetry and concentration, the KS test against the ACS field galaxy distributions fails to provide constraints, as we might expect from the broad range of C and A values shown in Table~\ref{tab:CAS}.

As another approach, we have taken a sample of local ULIRGs and normal galaxies and simulated how they would look in the GOODS ACS images at higher redshifts and then measured their CAS parameters~\citep{cc_CAS}. In this way, we directly compare the CAS parameters of the sub-mm sample to more specific high-redshift populations. We have simulated the appearance of these galaxies at both $z\simeq1$ and $z\simeq2$. We expect the sub-mm sample to look most like the ULIRGs at $z\simeq2$, since this is closer to the median redshift of the sub-mm galaxies. The right panel of Fig.~\ref{fig:petr} shows the distribution of the Petrosian radius for the sub-mm sample, the $z\simeq2$ ULIRG sample and the $z\simeq2$ normal galaxy sample. From this plot, the sub-mm galaxies can clearly be ruled out as being normal galaxies in terms of size, while they appear closer to the high-redshift ULIRG size distributions. The KS test rules out the sub-mm galaxies being like normal galaxies at both redshifts in terms of size and asymmetry to $>99$ per cent and to $>90$ per cent in terms of concentration. On the other hand, the hypothesis that the sub-mm galaxies are drawn from same population as the ULIRGs in terms of concentration, asymmetry and radius cannot be rejected.

\begin{figure}
\begin{center}
\includegraphics[width=3.0in,angle=0]{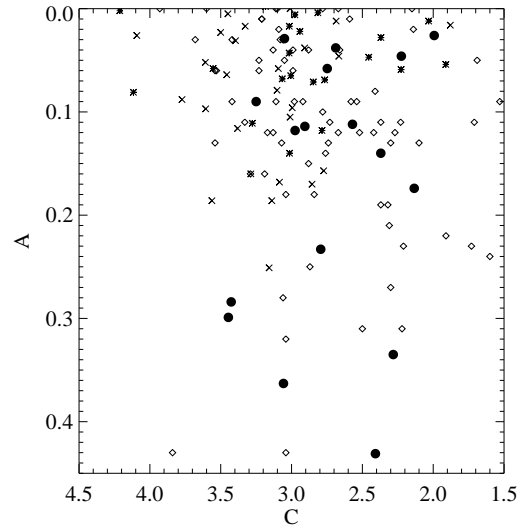}
\caption{Concentration versus Asymmetry for sub-mm sources and other high-redshift populations (based on photometric redshifts). The solid circles are our sub-mm sources and the hollow diamonds are high-redshift ($z>1.2$) EROs with $i_{\rm{775}}-K_{\rm{s}}>2.5$ and $J-K_{\rm{s}}>1.0$. The asterisks and the crosses show the simulated $z\simeq2$ ULIRGs and normal galaxies, respectively.
}
\label{fig:CA}
\end{center}
\end{figure}

Fig.~\ref{fig:CA} shows the concentration versus asymmetry plane. In the local Universe, galaxies of different types lie on different parts of this plot~\citep[see][]{cc_CAS}. The sub-mm galaxies are clearly mixed in terms of concentration and asymmetry, as are the other high-redshift populations on the plot. However, it may be noteworthy that the highest asymmetry points tend to be either sub-mm galaxies or high-redshift EROs.

\cite{Chapman_morph} completed a morphological study of sub-mm galaxies using \hst\ $R$-band images to a depth of $R(\rm{AB})\simeq27$. They found that about 60 per cent of the sample showed evidence for an active merger, and about 70 per cent were extremely large relative to the field population, regardless of optical magnitude~\citep{cc_submm}. The merger fraction is based on adjusting the sub-mm sources for redshift by adding a correction factor to the asymmetries of the SCUBA galaxies. However, this correction does not include any effects of the K-correction. 

Several of our sub-mm sources lie in the original HDF region. For these galaxies, GN13, GN28 and GN30, we measured the CAS parameters in the WFPC2 optical wavebands and in the deep NICMOS $J$ and $H$ images~\citep{Dickinson99,Dickinson00} in addition to the ACS bands. We found that the parameters that we are most interested in, namely radius and asymmetry, change little between the red optical bands and the near-IR bands. This is reassuring, because it means that the wavelength difference within the sub-mm sample itself will not have a major effect on the parameters.

\section{Conclusions}

Using the optical and near-IR images from GOODS and guided by the properties of the radio-detected galaxies, we have identified and characterized a large fraction of our sample of sub-mm sources. 

We have added to our statistically robust sub-mm source list in GOODS-North to give a total of 40 sources detected at $>3.5\sigma$ at 850$\,\mu$m, one of which is the brightest known `blank sky' extragalactic SCUBA source and is radio-unidentified. About $3/4$ of our sample has been identified with a unique optical or near-IR counterpart by applying new techniques for identifying optical counterparts using the properties of the radio-detected galaxies as a guide. This is the first time that plausible counterparts have been found for a significant fraction of radio-undetected sub-mm sources ($12/24$ unique counterparts). An additional 18 per cent of our sources have several possible counterparts that meet our criteria. Therefore only about 10 per cent of our sample have no counterparts in the GOODS images and thus our identification is close to complete. This means that for the first time we can be confident that the correct counterparts for SCUBA sources are unlikely to be invisible in the available optical images. 

With the deep optical and near-IR photometry, we estimate photometric redshifts for our sub-mm sample and find a median redshift of 2.0. When separated into radio-detected and radio-undetected sub-populations, the medians are 1.7 and 2.3, respectively. However, we are unable to show that the samples are drawn from different populations in redshift because of the small numbers.

Correlations between the 850$\,\mu$m flux and both the $i_{\rm{775}}$ and the redshift are intriguing in that there appears to be an absence of high redshift faint counterparts to the lower flux density SCUBA sources. We are currently unable to distinguish between several possible explanations for this, but evolution is likely to play a role. 

The sub-mm galaxies are red both in $i_{\rm{775}}-K_{\rm{s}}$ and $J-K_{\rm{s}}$. These colours together are most useful for selecting and describing the counterparts. The fraction of near-IR counterparts which are classified as EROs is much higher for the radio-detected sub-mm sources than for the radio-undetected sub-mm sources. This may indicate differences in their spectral energy distributions if they are at similar redshifts, and is another topic which needs to be investigated further.

Although sub-mm galaxies show a range of morphologies, in terms of concentration and asymmetry they are generally larger than field galaxies, consistent with being the most massive galaxies at all redshifts. By simulating normal galaxies shifted to the redshifts of the sub-mm sources, we can rule out the possibility that the high-redshift sub-mm galaxies could have normal morphologies.

The obvious next step is to extend the study of this sample into the infrared using \spitzer. As part of GOODS, observations with IRAC and MIPS of GOODS-North are ongoing and raw data are becoming available in the archive. Early results from \spitzer\ have shown that a high percentage of SCUBA sources are recovered (Egami et al.~2004; Ivison et al.~2004). We will use the colours in the \spitzer\ bands to verify counterparts and the additional photometry points will be invaluable in understanding the optical/IR SEDs of SCUBA galaxies. 

In the absence of a radio counterpart we will not be able to localize the sub-mm emission until the arrival of ALMA, or perhaps a fully operational SMA. However, in the next couple of years, SHADES, the largest survey so far conducted with SCUBA, is expected to reveal about 300 850$\,\mu$m sources. And not much further in the future, we have the prospect of samples of many thousands of sub-mm sources detected with SCUBA-2. At the moment, it is hard to detect each individual SCUBA source and the samples are quite modest in size. With the advent of  much larger surveys compiled in a systematic way, it should be possible to perform much more extensive and precise studies of counterparts to sub-mm galaxies. Our current work suggests that optical imaging to $i_{\rm{775}}\simeq28$, coupled with deep radio data, should be sufficient to identify the bulk of the sub-mm galaxies in these future surveys.

\section*{Acknowledgments}
We thank the referee, Rob Ivison, for his careful reading of this manuscript and useful suggestions. We are grateful to Mark Halpern for helpful comments, and particularly for suggestions for improving the extraction of adjacent sources. We would like to thank Kyoungsoo Lee for reducing the GOODS Flamingos data used in this paper. This work was supported by the Natural Sciences and Engineering Research Council of Canada. The James Clerk Maxwell Telescope is operated by The Joint Astronomy Center on behalf of the Particle Physics and Astronomy Research Council of the United Kingdom, the Netherlands Organisation for Scientific Research, and the National Research Council of Canada. Much of the data used for our analysis was obtained via the Canadian Astronomy Data Centre, which is operated by the Herzberg Institute of Astrophysics, National Research Council of Canada. Support for the GOODS HST program was provided by NASA through grant GO09583.01-96A from STScI, which is operated by AURA, Inc., under NASA contract NAS5-26555.



\appendix

\section{New submillimetre observations and source list}

\begin{figure*}
\begin{center}
\includegraphics[width=6.0in,angle=0]{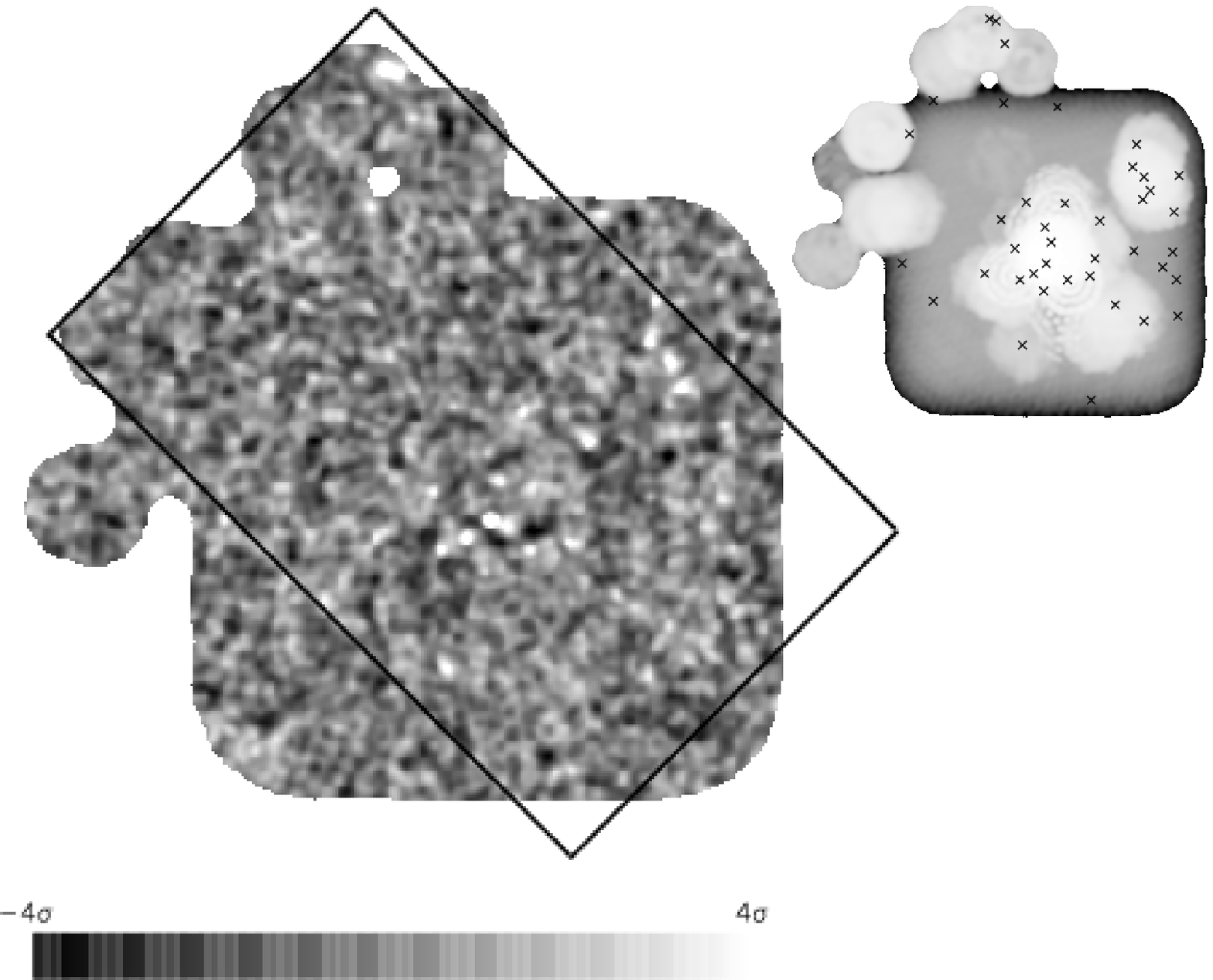}
\caption{The 850$\,\mu$m signal-to-noise super-map of the HDF-N region. The rectangle shows the boundary of the GOODS-North region. This map has been cleaned to remove the negative beams of the sources. The noise map is shown in the top right corner, with the crosses showing the positions of the 40 sources.
}
\label{fig:supermap}
\end{center}
\end{figure*}

\subsection{New 850$\,\mu$m observations}
The sub-mm observations presented in Paper I and Paper II contain all the data taken by our group up to Fall 2002 and all other data publicly available in the CADC JCMT archive at that time. Since then our group has collected an additional 32 hours of SCUBA jiggle-map data and 4 hours of SCUBA photometry in GOODS North. More sub-mm data from other groups within the region have also become available in the archive. All of the new data have been combined with the Paper I data to make an updated super-map of GOODS North. 

Most of our new sub-mm data was taken in the SCUBA jiggle-map mode, where we have adopted the same multi-chopping strategy as that of SHADES~\citep{Mortier04}. Each new target position was observed with 6 different chop configurations, namely chop throws of $30\,$arcsec, $44\,$arcsec and $68\,$arcsec at chop angles of both 0 degrees and 90 degrees. This reduces the risk of missing (or finding false) sources because they were chopped onto, or because of bad bolometers. It also allows us to test the robustness of sources when we detect the negative beams in each of the sub-maps.

Between December 2002 to June 2003, a peculiar noise spike was found in the power spectrum of some of the bolometers in the sub-mm data. This noise occurs on the same timescale as every 16 samples, which matches the number of samples per quadrant of the 64 point jiggle scheme and therefore introduces difficulties for sky subtraction. Although there has been some investigation, there is currently no known solution for this problem. We checked all of our new observation files and only a very small fraction contain the noise spike. We have opted to remove these files from the final map altogether, in order to ensure that no false structure is introduced. 

The JCMT has also recently reported an error in the tracking model, which affected the pointing accuracy of some data files from August 2000 to April 2003, depending on the elevation~\citep{Tilanus04}. This error is expected to cause a maximum shift of $5\,$arcsec. We found that 4 of the archive files which were included in the super-map are affected by this tracking error and we have removed these files from the map. This has not affected any of the sub-mm sources in our catalogue.

We follow the data reduction and source extraction procedure as described in Paper I~\citep[see also][]{cb_thesis} with a few minor improvements. Starting in 2003, the JCMT Water Vapor Monitor (WVM, B. Weferling, private communication) became the primary device used for atmospheric corrections, since it provides much more frequent measurements. In our new observations, we use the WVM to correct for sky extinction~\citep{Archibald02} whenever those data are available. We have also updated the calibration factors used for all observation files since 1997. We calculated the flux conversion factors (FCFs) for a given observation from the average of the calibrations from the same night taken in the same SCUBA mode. These calculated values are used for calibration unless they show a large variance, in which case we use the new standard FCFs published on the JCMT web-page\footnote{http://www.jach.hawaii.edu/JACpublic/JCMT/Continuum\_observing\-/SCUBA/astronomy/calibration/gains.html}, which are tabulated approximately monthly.

The noise in the super-map is very non-uniform, due to combining all different modes of SCUBA observations (scan-mapping, jiggle-mapping and photometry), as well as different exposure times and different chopping configurations. The new super-map covers a total area of approximately 200 square arcminutes with an average $1\sigma$ RMS of $3.4\,$mJy. However, half of the super-map is much deeper with a $1\sigma$ RMS of $<2.5\,$mJy, and 70 per cent of our sources come from this deeper region. Fig.~\ref{fig:supermap} shows the SCUBA signal-to-noise super-map along with the noise map which also indicates the locations of the sources.

\subsection{New 850$\,\mu$m source list}

We detect 21 sources at $>4\sigma$ in the 850$\,\mu$m map and an additional 17 sources if we consider $>3.5\sigma$ (and 2 other companion sources appear after a `clean' process which we discuss in Section A3). Table~\ref{tab:cat850} gives the new source list, including those from Papers I and II, with the updated positions and fluxes. The GN identification number is used when referring to the sources in this paper and full SMM name is listed for future reference to these sources.

All of the $>4\sigma$ sources from Paper II were recovered in the new reduction, with some minor changes in position and flux values. One of the Paper II sources from the supplementary $3.5$--$4.0\sigma$ list (SMMJ123719+621107) is no longer detected at $>3.5\sigma$. This is not surprising, since the Monte Carlo simulations performed in Paper I tell us that we can expect that on average there will be 2 spurious sources in the supplementary catalogue. 

There are 7 new sources in the 850$\,\mu$m super-map,  3 in the $>4\sigma$ list and 4 in the $3.5$--$4.0\sigma$ list. Of these, 4 are from the new data collected in 2003. The most interesting of these sources is an unusually bright system with a total 850$\,\mu$m flux of around 30$\,$mJy, in a region with a noise level of only 2$\,$mJy. This is unprecedented for `blank sky' SCUBA sources. When we reduce the data for the different chop configurations separately, this source and the negative beams are obvious in each case, which confirms that the source is not spurious. This new bright source is discussed in Section A4.

\subsection{A `cleaned' 850$\,\mu$m map}

Although the negative sidelobes of a given source help to identify it as such, they can also interfere with other sources nearby. Since the super-map contains SCUBA data collected with different chop patterns, it is difficult to distinguish off-beams when examining the super-map alone. In order to overcome this problem, we adopted the following procedure. First, we constructed the super-map and extracted sources, as in Paper I. We then re-reduced the data, starting with a model consisting of point sources included at the level of the time stream differences. This resulted in a map which was effectively cleaned of both the positive signal and the negative sidelobes of all detected sources. We then re-ran the source extraction algorithm to find any new sources. And finally we iterated the whole process several times to check that the source list was stable. 

We found 2 new sources after cleaning the 850$\,\mu$m super-map, and 2 more are just below our $3.5\sigma$ threshold. Both new sources are within 30 arcsec of one of the initial sources, and therefore we call each a `companion source'. We derive the best estimate for the flux of both the initial source and the companion by fitting to the data two Gaussians with variable positions and amplitudes. We found that the fluxes and positions could only be reliably estimated through a simultaneous fitting procedure.

These new detections may be the result of sources being extended, but this is unlikely given that 20 arcsec corresponds to $\sim150\,$kpc at $z=3$. We can also rule out the possibility that these detections are a product of a non-circular Gaussian PSF, since, at 15 arcsec from the peak of a source, the difference between a typical SCUBA beam shape and a Gaussian could only give rise to a $3.5\sigma$ detection if the initial source has a signal-to-noise ratio of at least 18. Given that the other explanations seem unlikely, we treat these as new sources and list them along with the others in Table~\ref{tab:cat850}. The companion sources are listed in the table with IDs that indicate which initial source they are close to and their possible counterparts are discussed in the next section. 
  
Since the negative sidelobes do not represent real features in GOODS-North, we have made our final super-map by re-inserting the sources, without their negative signature, into the cleaned map. This map (Fig.~\ref{fig:supermap}) is an honest representation of the sub-mm data in the region. The beam patterns of all the $>3.5\sigma$ sources have been removed, although the map still contains the negative beams of the $<3.5\sigma$ sources. Comparison of Fig.~\ref{fig:supermap} with fig. 5 of Paper I shows the improvement. Due to the differential method in which SCUBA collects data, there is no sensitivity to the overall DC level in any image, and we expect the mean of a SCUBA map to be consistent with zero. This is in fact the case for both the initial signal map and the cleaned map without the sources. However, because we are only adding the positive contribution of our sources back into the cleaned map, this final cleaned map with sources will have a slightly positive mean, indicating the presence of these sources.

\begin{table*}  
\caption{Updated list of 850$\,\mu$m sources in the GOODS-North region. The table lists the 21 sources detected at $>4\sigma$ followed by 17 sources detected initially at $3.5$--$4.0\sigma$. The 2 extra sources found by cleaning our map are listed at the bottom of the table. For convenience we list the sources according to a GOODS-North (GN) ID number. Sources that are new since Paper II are listed as such in the Comments column. The last column lists the status of the optical counterpart for each sub-mm source as: secure (S); tentative (T); multiple possible counterparts that meet our counterpart selection criteria (M); or no counterpart (N). There are several minor changes compared with the table 4 of Paper II, which we discuss in the text. Note that GN38 is just outside the GOODS-North region. 
}
\normalsize
\label{tab:cat850}
\begin{tabular}{llllllll}
\hline
SMM ID  &  SMM Name  & RA     & DEC    & $S_{\rm 850\,\mu m}$(mJy) & SNR  & Comments & Optical ID\\\hline\hline
GN01   & SMMJ123606+621556 & 12:36:06.7 & 62:15:56 & $7.3\pm1.5$   & 4.9  &  New & T \\
GN02   & SMMJ123607+621147 & 12:36:07.7 & 62:11:47 & $16.2\pm4.1$  & 4.0  &  & M \\
GN03   & SMMJ123608+621253 & 12:36:08.9 & 62:12:53 & $16.8\pm4.0$ & 4.2  &  & T \\
GN04   & SMMJ123616+621520 & 12:36:16.6 & 62:15:20 & $5.1\pm1.0$   & 5.1  &  & S  \\
GN05   & SMMJ123618+621008 & 12:36:18.8 & 62:10:08 & $6.7\pm1.6$   & 4.2  &  & T  \\
GN06   & SMMJ123618+621553 & 12:36:18.7 & 62:15:53 & $7.5\pm0.9$   & 8.3  &  & S  \\
GN07   & SMMJ123621+621711 & 12:36:21.3 & 62:17:11 & $8.9\pm1.5$   & 5.9  &  & S  \\
GN08   & SMMJ123622+621256 & 12:36:22.2 & 62:12:56 & $12.5\pm2.7$  & 4.6  &  & T \\
GN09   & SMMJ123622+621617 & 12:36:22.6 & 62:16:17 & $8.9\pm1.0$   & 8.9  &  & M \\
GN10   & SMMJ123633+621408 & 12:36:33.8 & 62:14:08 & $11.3\pm1.6$  & 7.1  &  & T \\
GN11   & SMMJ123637+621156 & 12:36:37.2 & 62:11:56 & $7.0\pm0.9$   & 7.8  &  & S  \\
GN12   & SMMJ123645+621450 & 12:36:45.8 & 62:14:50 & $8.6\pm1.4$   & 6.1  &  & S  \\
GN13   & SMMJ123650+621317 & 12:36:50.5 & 62:13:17 & $1.9\pm0.4$   & 4.8  &  & S  \\
GN14   & SMMJ123652+621226 & 12:36:52.2 & 62:12:26 & $5.9\pm0.3$   & 19.7  & HDF850.1$^{\rm{a}}$  & S \\
GN15   & SMMJ123656+621202 & 12:36:56.5 & 62:12:02 & $3.7\pm0.4$   & 9.3  &  & T \\
GN16   & SMMJ123700+620911 & 12:37:00.4 & 62:09:11 & $9.0\pm2.1$   & 4.3  &  & S  \\
GN17   & SMMJ123701+621147 & 12:37:01.2 & 62:11:47 & $3.9\pm0.7$   & 5.6  &  & S  \\
GN18   & SMMJ123703+621302 & 12:37:03.0 & 62:13:02 & $3.2\pm0.6$   & 5.3  &  & M \\
GN19   & SMMJ123707+621411 & 12:37:07.7 & 62:14:11 & $10.7\pm2.7$  & 4.0  &  & S  \\
GN20   & SMMJ123711+622212 & 12:37:11.7 & 62:22:12 & $20.3\pm2.1$  & 9.7  & New  & T \\
GN21   & SMMJ123713+621202 & 12:37:13.3 & 62:12:02 & $5.7\pm1.2$   & 4.8  &   & N \\\hline
GN22   & SMMJ123607+621020 & 12:36:07.3 & 62:10:20 & $14.4\pm3.9$  & 3.7  &  & S \\
GN23   & SMMJ123608+621429 & 12:36:08.4 & 62:14:29 & $7.0\pm1.9$   & 3.7  &  & S  \\
GN24   & SMMJ123612+621217 & 12:36:12.4 & 62:12:17 & $13.7\pm3.6$  & 3.8  & & T \\
GN25   & SMMJ123628+621047 & 12:36:28.7 & 62:10:47 & $4.6\pm1.2$   & 3.8  & & S \\
GN26   & SMMJ123635+621238 & 12:36:35.5 & 62:12:38 & $3.0\pm0.8$   & 3.8  & & S \\
GN27   & SMMJ123636+620659 & 12:36:36.9 & 62:06:59 & $24.0\pm6.1$  & 3.9  & & M \\
GN28   & SMMJ123645+621147 & 12:36:45.0 & 62:11:47 & $1.7\pm0.4$   & 3.8  &  New & M \\
GN29   & SMMJ123648+621841 & 12:36:48.3 & 62:18:41 & $20.4\pm5.7$  & 3.6  & & M \\
GN30   & SMMJ123652+621353 & 12:36:52.7 & 62:13:53 & $1.8\pm0.5$   & 3.6  & & S \\
GN31   & SMMJ123653+621120 & 12:36:53.1 & 62:11:20 & $2.8\pm0.8$   & 3.5  & & T \\
GN32   & SMMJ123659+621453 & 12:36:59.1 & 62:14:53 & $5.3\pm1.4$   & 3.8  & & T \\
GN33   & SMMJ123706+621850 & 12:37:06.9 & 62:18:50 & $21.7\pm5.8$  & 3.7  & & N \\
GN34   & SMMJ123706+622112 & 12:37:06.5 & 62:21:12 & $5.6\pm1.6$   & 3.5  &  New & T \\
GN35   & SMMJ123730+621056 & 12:37:30.8 & 62:10:56 & $14.3\pm3.9$  & 3.7  &  & N \\
GN36   & SMMJ123731+621856 & 12:37:31.0 & 62:18:56 & $24.8\pm7.0$  & 3.5  &  & N \\
GN37   & SMMJ123739+621736 & 12:37:39.1 & 62:17:36 & $6.8\pm1.9$   & 3.6  &  New & T \\
GN38   & SMMJ123741+621226 & 12:37:41.6 & 62:12:26 & $24.9\pm6.5$  & 3.8  & No ACS & N \\\hline\hline
GN04.2 & SMMJ123619+621459 & 12:36:19.2 & 62:14:59 & $3.6\pm1.0$   & 3.6  & New, clean  & M \\
GN20.2 & SMMJ123709+622206 & 12:37:09.5 & 62:22:06 & $11.7\pm2.2$  & 5.3 & New, clean & S \\\hline
\end{tabular}
\medskip
\\
$^{\rm{a}}$\,See \cite{Dunlop04} for detailed counterpart analysis of this source. Our `cleaning' procedure finds a companion source about 25 arcsec to the south-west. However a simultaneous fit places it at a SNR of slightly less than 3.5.\\
\end{table*}

\subsection{New 850$\,\mu$m sources}
Each of the initial super-map sources is discussed in detail in Paper II. Here we give a brief description of the multi-wavelength environments of the new sources. Designations, coordinates and fluxes are listed in Table~\ref{tab:cat850}.

\noindent\textbf{GN01:} This source is also detected in the ~\cite{Wang04} HDF sub-mm study at a similar flux level. There are six optical galaxies within our $7\,$arcsec search radius. However, only one is detected in the near-IR. This near-IR source is also detected in the X-ray and there is a hint of a detection in the radio at $<5\sigma$. We identify this galaxy with the sub-mm emission. It has a spectroscopic redshift of 2.4 from~\cite{Cowie04}.

\noindent\textbf{GN20:} This source is one of the brightest blank-sky sub-mm sources ever detected with SCUBA and is a very convincing example of a radio-undetected SCUBA source. Given its signal-to-noise ratio, the fact that it is separately detected in all 6 chop-throw sub-maps, and the fact that our `poorness-of-fit' statistics (see Appendix B) find nothing unusual at its position, then the chance that this source is spurious must be extremely low. It is also so far above the confusion limit that confusion and flux-boosting cannot play a significant role. This new source is not detected in the 450$\,\mu$m map (although the RMS there is rather poor), nor is it detected in the deep VLA radio images ($1\sigma$ RMS of $\sim10\,\mu$Jy) or the Chandra $2\,$Msec image.  Using the far-IR/radio correlation, these non-detections imply that the sub-mm bright galaxy is at a redshift of at least 2. The ACS images reveal 13 optical galaxies within the $7\,$arcsec search radius, 2 of which are detected above $5\sigma$ in the $K$-band image. If we consider the redshift constraints from the lack of radio flux, then we can eliminate one of these, and our identification technique selects the other red source. We have recently acquired follow-up continuum mapping at $1.3\,$mm with the IRAM Plateau de Bureau interferometer to localize the sub-mm emission and confirm the correct optical counterpart. A more detailed study of this unique source is reserved for a future paper (Pope et al., in preparation).

While it has been argued that many RU sub-mm sources may be spurious, this source is clearly very robust. Given how many of the radio identifications are close to the detection limit, it is not surprising that such sources exist; it simply implies that some fraction of sub-mm sources are undetected in the radio and are at $z\ge2$ and/or have more unusual SEDs than we expect.

\noindent\textbf{GN28:} This source resides in a crowded area surrounded by 11 optical sources and 1 X-ray source. The X-ray source is not coincident with any of the $>5\sigma$ optical sources, although there is a faint smudge in the $K$-band image. None of the optical galaxies stand out with our identification criterion (see Section 5). Therefore, while it is likely that the sub-mm emission is associated with the obscured X-ray source, we list this source as having multiple possible optical counterparts.

\noindent\textbf{GN34:} There are at least 3 bright optical galaxies with photometric redshifts of around 1 within $7\,$arcsec of this SCUBA source. Our counterpart selection technique clearly selects an ERO with red near-IR colours as the most likely counterpart.

\noindent\textbf{GN37:} Only 2 of the 14 optical galaxies within $7\,$arcsec of this sub-mm source have any near-IR emission. One is very bright and has a secure photometric redshift of 0.43, while the other has a spectroscopic redshift of 3.2~\citep{Cowie04}. The first source can be eliminated due to the lack of radio emission and we tentatively identify this source with the red $z\simeq3.2$ galaxy.

\noindent\textbf{GN04.2:} This source is also detected in~\cite{Wang04} at an identical flux level, although there is no obvious optical counterpart. There are many possible optical sources, including a Lyman break galaxy, however none of them meet our selection criteria.

\noindent\textbf{GN20.2:} This source is the companion to our new bright source. It is $20\,$arcsec away, which implies that it is a distinct object. However, we should also consider the possibility that GN20 is extended (or has multiple images) due to lensing. GN20.2 does have a radio source nearby and therefore, given our criteria for identification, this is the secure counterpart. It has a photometric redshift of 3.91, and is very well fit to a starburst template. 

\subsection{450$\,\mu$m observations and source list}
While the focus is normally on the 850$\,\mu$m observations in extragalactic sub-mm surveys, SCUBA simultaneously collects data at both 450$\,\mu$m and 850$\,\mu$m. The atmosphere in the shorter waveband is much worse than at 850$\,\mu$m and therefore observing conditions that are adequate for 850$\,\mu$m observations are not sufficient to provide high quality 450$\,\mu$m data. Nevertheless, we have reduced the 450$\,\mu$m data and compiled a new source list.

Since Paper II, the JCMT announced that two of the bolometers in the 450$\,\mu$m array experience cross-talk, and therefore should be removed from any data taken since 1997\footnote{http://www.jach.hawaii.edu/JACpublic/JCMT/Continuum\_observing\-/SCUBA/news/message.html}. These bolometers could easily introduce false structure and sources in the 450$\,\mu$m super-map, and therefore we do not use any of the data from them.

There are 7 sources detected at $>4\sigma$ in our 450$\,\mu$m super-map (Table~\ref{tab:cat450}), and no sources detected at $>5\sigma$. Paper II reported 5 detections at this wavelength, but only one of these is in our new $>4\sigma$ list. 2 others from Paper II (SMMJ123632+621542 and SMMJ123727+621042) are detected at $3.5$--$4.0\sigma$ in the new map, but we stick with $>4\sigma$ detections at 450$\,\mu$m, since the data are already so noisy. The remaining Paper II 450$\,\mu$m detections are not detected in the new map down to $3.5\sigma$. Given the two bad bolometers we have removed in addition to the poorly behaved noise (still partially correlated, since the sky is not accurately removed) and our new calibration procedure, this low detection rate of Paper II sources is not surprising.

It is worth noting that none of the $>3.5\sigma$ 850$\,\mu$m sources match up with $>3.5\sigma$ 450$\,\mu$m sources. However, HDF850.1 is the closest, with a $3.5\sigma$ 450$\,\mu$m source $12\,$arcsec from the 850$\,\mu$m position.

There are no radio or X-ray sources in the vicinity of any of the 450$\,\mu$m sources. Each source has from 0 to 7 optical sources within the search radius, but nothing stands out as a counterpart. The 450$\,\mu$m data in this survey suffer from not having the very best weather. Most of the super-map observations were completed in Grade 2 and 3 weather, while high-quality 450$\,\mu$m data are achieved only in Grade 1. In good weather, 450$\,\mu$m photometry would be able to detect most of the radio-detected 850$\,\mu$m sources and help with SED constraints. However, with the current quality of data, no 850$\,\mu$m source is convincingly detected, although the stacked 450$\,\mu$m flux at the positions of the the 850$\,\mu$m sources is mildly detected at $(4.9\pm1.8)\,$mJy. For comparison, the stacked 850$\,\mu$m flux of the 450$\,\mu$m sources is $(-2.0\pm0.8)\,$mJy. 

A detection of negative sources, which we know must be spurious, would indicate poor reliability in the positive sources at either wavelength. In the 850$\,\mu$m map, we detect 4 negative sources not associated with the off-beams of a positive source, which is within the error bars of the number of spurious sources we expect from the Monte Carlo simulations (see Paper I), whereas in the 450$\,\mu$m map, there are 9 negative detections not associated with a positive source. Given these statistics and the only modest overlap with the 450$\,\mu$m sources in Paper II, we cannot be confident of the reality of {\it any} of the 7 new candidate 450$\,\mu$m sources. This is in contrast with the situation at 850$\,\mu$m, where there is extremely good agreement with the source list in Paper II.

\begin{table*} 
\caption{450$\,\mu$m source list. There are 7 candidate sources detected at $>4\sigma$, however none of them are detected at $>5\sigma$. We list the $3\sigma$ upper limits at 850$\,\mu$m and also give comments on some sources. }
\normalsize
\label{tab:cat450}
\begin{tabular}{llllll}
\hline
SMM ID  & RA             & DEC           & $S_{\rm 450\mu m}$  & $S_{\rm 850\mu m}$ & Comment  \\
                       &                &               & (mJy)     & (mJy)        &  \\\hline\hline
SMMJ123603+620942 & 12:36:03.0  & 62:09:42  &	$193\pm49$  & $<14.4$   &  \\
SMMJ123631+620657 & 12:36:31.8  & 62:06:57  &	$269\pm65$  & $<18.6$   &   \\
SMMJ123638+621012 & 12:36:38.6  & 62:10:12  &	$77\pm19$   & $<3.6$   &  \\
SMMJ123649+620918 & 12:36:49.7  & 62:09:18  &       $141\pm29$  & $<6.6$   &   \\
SMMJ123657+622033 & 12:36:57.9  & 62:20:33  &	$223\pm55$  & $<7.8$   &    New data in this region \\
SMMJ123702+621012 & 12:37:02.6  & 62:10:12  &	$111\pm25$  & $<5.1$   &  Paper II \\
SMMJ123747+621560 & 12:37:47.3  & 62:15:60  &	$291\pm71$  & $<10.2$   &  New data in this region \\

\end{tabular}
\end{table*}

\section{Robustness of our sub-mm sample}

\begin{figure*}
\begin{center}
\includegraphics[width=6.0in,angle=0]{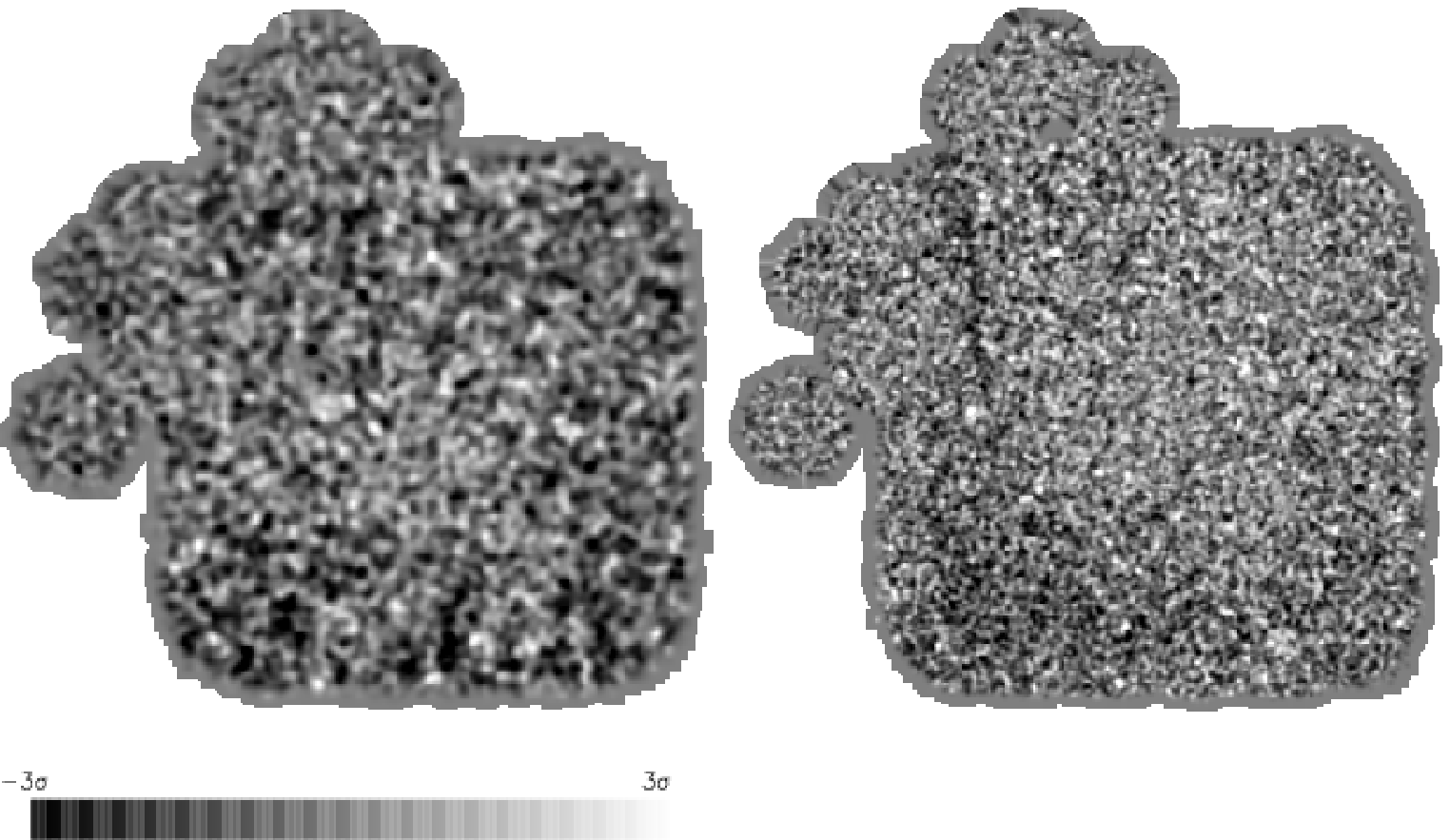}
\caption{Smoothed `temporal' $\chi^{2}$ SNR maps at 850$\,\mu$m (left) and 450$\,\mu$m (right). By using these maps, we can pick our regions of the super-map with significantly bad $\chi^{2}$ fits (i.e. internal inconsistencies of the data contributing to each pixel in the map). For example, both the 850$\,\mu$m and 450$\,\mu$m maps show a pattern of bad pixels in the bottom left-hand corner of the square scan-map region. None of our sources lie in this area. However, if they did we would flag them as having potentially inconsistent data and would then investigate the source data at the timestream level. It is also clear that the 450$\,\mu$m $\chi^{2}$ map contains much more structure than the 850$\,\mu$m map. In general we find no coincidence between peaks in these maps and our sources (Fig.~A1) 
}
\label{fig:chi2_map}
\end{center}
\end{figure*}

The robustness of radio-undetected SCUBA sources has recently been challenged in the literature~\citep{Greve04}. But as discussed there, the lack of radio flux could have other reasonable explanations, in particular these sources could be at higher redshift, or have cooler temperatures, and therefore not be accessible with current radio telescopes. If the RU SCUBA sources are {\it all} spurious then this implies we do not fully understand the behaviour of the noise, since Monte Carlo simulations including realistic noise (see Paper I) lead us to expect only 2.5 spurious positive sources in our survey area. If this were the case, there should be some indication in the statistics of the raw data. Therefore we have performed several $\chi^{2}$ tests to determine how well the raw data fit with the final maps.

The first is a spatial $\chi^{2}$ test, which provides a measure of how well the Gaussian PSF fits the data at each source position in the final super-map. If there are a significant number of pixels that disagree with the best-fit source, it will be shown in this spatial $\chi^{2}$ fit. This test indicates whether a source is extended, blended, confused or otherwise a poor fit to a point source.

We also performed a perhaps more useful temporal $\chi^{2}$ test, which we now describe. Our reduction code creates the signal map by performing a weighted average of all data that hit a certain pixel, where the noise for each pixel comes from the weighted variances of the bolometers (see Paper I for more details). For a given position in the final map, it is useful to know the overall consistency of the raw data that contributed to that pixel. This would tell us if, for example, half of the hits on a certain pixel are consistent with one value, while the other half prefer another value -- in that case our map would suggest an average value for the pixel, which nevertheless might be a poor fit to any of the data. The temporal $\chi^{2}$ assesses this self-consistency of the data.

\begin{figure}
\begin{center}
\includegraphics[width=3.0in,angle=0]{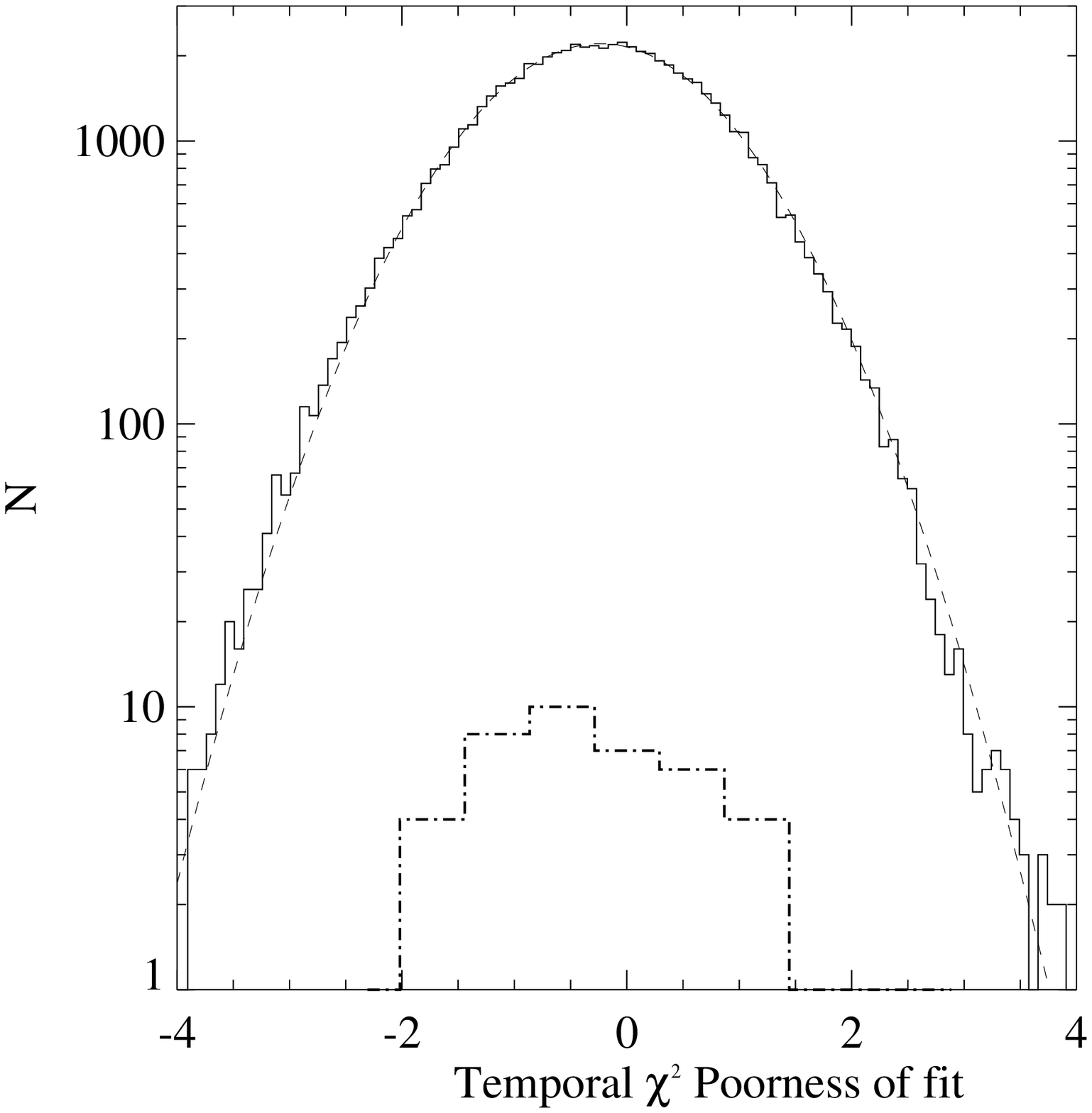}
\caption{Distribution of `Poorness of fit' for the temporal $\chi^{2}$ statistic at 850$\,\mu$m (i.e. the $\chi^{2}$ for the fit of the raw data to the final map, converted into an approximately Gaussian variate). The solid curve at the top is the distribution of all pixels in the map with at least 100 hits. The dashed line is the best-fit Gaussian. The fit is very close to a normal distribution with the mean just below 0 (which implies that we are slightly underestimating our noise) and a variance of close to 1. The lower histogram is the distribution for all 40 sub-mm sources. This sub-mm distribution is completely consistent with that of the map as a whole, with none of the sources standing out as being a very poor fit.
}
\label{fig:chi2_dist}
\end{center}
\end{figure}

We calculated such a $\chi^{2}$ map for our $3\times3$ arcsec pixels, relative to the final super-map as the model. We then investigated whether outliers are correlated with the positions of any of our sources, or in other words whether some of our sources may be in areas of `bad' data. However, since each source actually contains a contribution from roughly a beam-area of pixels, we need to consider a smoothed version of this $\chi^{2}$ map. So to assess whether a source has a poor temporal $\chi^{2}$, we can proceed as follows. First we calculate the pixel temporal $\chi^{2}_{i}$ and number of hits for each pixel, $N^{i}_{\rm{hits}}$. We then expect an average of $N^{i}_{\rm{hits}}$ for pixel $i$, and a variance of $2N^{i}_{\rm{hits}}$. For large enough values of $N^{i}_{\rm{hits}}$, the quantity $\left(\chi^{2}_{i}-N^{i}_{\rm{hits}}\right)/\sqrt{2N^{i}_{\rm{hits}}}$ should be approximately Gaussian distributed with mean zero and variance 1. So in the same way that the signal-to-noise super-map gives the best estimate of the SNR of a point source centred on each pixel, we can calculate a signal-to-noise map of $\chi^{2}$ (i.e. poorness of fit to the model) by treating the $\chi^{2}_{i}-N^{i}_{\rm{hits}}$ as the signal and $\sqrt{2N^{i}_{\rm{hits}}}$ as the noise, and performing a weighted convolution with the PSF in exactly the same way as we did for the super-map itself. The result is shown in Fig.~\ref{fig:chi2_map} at both 850$\,\mu$m and 450$\,\mu$m. We can see that there are few points of high `poorness of fit SNR' at 850$\,\mu$m, while the data are less well behaved at 450$\,\mu$m, as expected. Fig.~\ref{fig:chi2_dist} shows the distribution of pixel values in the 850$\,\mu$m map , with the values at the positions of our sources also shown for comparison.

We calculated several variations of both the spatial and temporal $\chi^{2}$ tests for all the sources in our catalogue. The spatial PSF fits are within $\pm2\sigma$ for all but two of our initial 38 sources, which is consistent with what we expect from Gaussian statistics. Of these two, one has a secure identification with a radio source, while the other is not detected in the radio. The temporal fits are within $\pm2\sigma$ for all the super-map sources (see Fig.~\ref{fig:chi2_dist}). There is no trend indicating that the values from the spatial or temporal fits get better, or worse, with noise level, flux level, or signal-to-noise ratio. There is also {\it no} discernible difference between the distributions for the RD and RU sources. When tested on the un-cleaned map, the 2 sources, which we now suggest are pairs of nearby sources, have much worse spatial fits, as we expect. However, their temporal fits are within the acceptable range and their spatial fits in the cleaned map are also reasonable.

Monte Carlo simulations discussed in Paper I (increased by about 20 per cent because of the increased survey area) tell us that we can expect an average of about 2.5 of the $3.5$--$4.0\sigma$ detected sources to be spurious. We performed tests to determine a reasonable threshold for both the spatial and temporal $\chi^{2}$ tests. Of our 40 sources, including the 2 `clean' sources, none of them show $\chi^{2}$ values significant at the $>3\sigma$ level. Based on these tests, we could find no reason to exclude any other sources. Hence, from investigating the statistics of the sub-mm map, there is no reason not to trust the radio-undetected sources.

We also performed all these tests on our 450$\,\mu$m data, and found that none of the 7 candidate sources were below the threshold. Although we might expect the 450$\,\mu$m fits to be worse than the 850$\,\mu$m fits (and despite the $\chi^{2}$ clearly showing some structure), these tests show that the data for the sources appear to be largely self-consistent.

\end{document}